\documentclass{aa}  
\usepackage{graphicx}
\usepackage{txfonts}
\usepackage{xcolor}
\usepackage{hyperref}
\hypersetup{
  colorlinks   = true, 
  urlcolor     = blue,
  linkcolor    = blue, 
  citecolor   = blue 
}

\usepackage{gensymb}
\usepackage{rotating}
\begin{document} 
\titlerunning{Spiral structure in the gas disc of CQ Tau}
\authorrunning{Wölfer et al.}

   \subtitle{Spiral structure in the gas disc of CQ Tau}

   \title{A highly non-Keplerian protoplanetary disc}

   \author{L. W\"olfer
          \inst{1}\fnmsep\inst{2},
           S. Facchini\inst{3},
           N.T. Kurtovic\inst{4}\fnmsep\inst{5},
           R. Teague\inst{6},
           E.F. van Dishoeck\inst{1}\fnmsep\inst{7},
           M. Benisty\inst{4}\fnmsep\inst{8}\fnmsep\inst{9},
           B. Ercolano\inst{2}\fnmsep\inst{10},
           G. Lodato\inst{11},
           A. Miotello\inst{3},
           G. Rosotti\inst{7},
           L.Testi\inst{3}\fnmsep\inst{10}\fnmsep\inst{12},
           M.G. Ubeira Gabellini\inst{3}\fnmsep\inst{11} 
          }

   \institute{Max-Planck-Institut f\"ur extraterrestrische Physik, Gie\ss enbachstr. 1 , 85748 Garching bei M\"unchen, Germany
   \\ e-mail: \href{mailto:woelfer@mpe.mpg.de}{woelfer@mpe.mpg.de}
         \and
             Universit\"ats-Sternwarte M\"unchen, Scheinerstr. 1, 81679 M\"unchen, Germany
        \and European Southern Observatory, Karl-Schwarzschild-Str. 2, 85748 Garching bei M\"unchen, Germany.
        \and Departamento de Astronom\'{i}a, Universidad de Chile, Camino El Observatorio 1515, Las Condes, Santiago, Chile
        \and Max-Planck-Institut f\"ur Astronomie, K\"onigstuhl 17, 69117 Heidelberg, Germany
        \and Center for Astrophysics $\vert$ Harvard \& Smithsonian, 60 Garden Street, Cambridge, MA 02138, USA
        \and Leiden Observatory, Leiden University, P.O. Box 9513, NL-2300 RA Leiden, The Netherlands
        \and Unidad Mixta Internacional Franco-Chilena de Astronom\`{i}a, CNRS, UMI 3386
        \and Univ. Grenoble Alpes, CNRS, IPAG, 38000 Grenoble, France
        \and Excellence Cluster Origin and Structure of the Universe, Boltzmannstr. 2, 85748 Garching bei M\"unchen, Germany
        \and Dipartimento di Fisica, Universit\`{a} degli Studi di Milano, Via Giovanni Celoria, 16, 20133 Milano, Italy
        \and INAF -- Osservatorio Astrofisico di Arcetri, Largo E. Fermi 5, 50125 Firenze, Italy 
             }

   \date{Received ; accepted}

 
  \abstract
   {In the past years, high angular resolution observations have revealed that circumstellar discs appear in a variety of shapes with diverse substructures being ubiquitous. This has given rise to the question of whether these substructures are triggered by planet-disc interactions. Besides direct imaging, one of the most promising methods to distinguish between different disc shaping mechanisms is to study the kinematics of the gas disc. In particular, the deviations of the rotation profile from Keplerian velocity can be used to probe perturbations in the gas pressure profile that may be caused by embedded (proto-) planets.}
   {In this paper we aim to analyze the gas brightness temperature and kinematics of the transitional disc around the intermediate mass star CQ Tau in order to resolve and characterize substructure in the gas, caused by possible perturbers.}
   {For our analysis we use spatially resolved ALMA observations of the three CO isotopologues $^{12}$CO, $^{13}$CO and C$^{18}$O $(J=2-1)$ from the disc around CQ Tau. We further extract robust line centroids for each channel map and fit a number of Keplerian disc models to the velocity field.}
   {The gas kinematics of the CQ Tau disc present non-Keplerian features, showing bent and twisted iso-velocity curves in $^{12}$CO and $^{13}$CO. Significant spiral structures are detected between $\sim$ 10-180\,au in both the brightness temperature and the rotation velocity of $^{12}$CO after subtraction of an azimuthally symmetric model, which may be tracing planet-disc interactions with an embedded planet or low-mass companion. We identify three spirals, two in the brightness temperature and one in the velocity residuals, spanning a large azimuth and radial extent. The brightness temperature spirals are morphologically connected to spirals observed in NIR scattered light in the same disc, indicating a common origin. Together with the observed large dust and gas cavity, these spiral structures support the hypothesis of a massive embedded companion in the CQ Tau disc.}
   {}


   \maketitle
%
\section{Introduction}
Due to angular momentum conservation, circumstellar discs are the natural outcome of the star formation process when infalling material from a molecular cloud core is channeled towards the newly formed central star. These accretion discs composed of gas and dust represent the nurseries of planetary systems. They evolve and ultimately disperse while giving birth to various objects with the evolutionary processes significantly influencing the ongoing planet formation. At the same time, the forming planets will backreact on the disc and affect its evolution and structure, resulting in a highly coupled and complex problem. 

High angular resolution observations indeed show that circumstellar discs are commonly marked by a variety of substructures in the gas and especially the dust, such as gaps, rings or even cavities, as well as spiral arms and azimuthal asymmetries (e.g. \citealp{Marel2013,Casassus2016,Andrews2018,Cazzoletti2018,Feng2018,Andrews2020}). Such substructures might be caused by embedded (proto-) planets (e.g. \citealp{Lin1986,Zhang2018,Lodato2019}), suggesting that planet formation occurs already in early evolutionary stages. However, there exist other mechanisms such as the magnetorotational instability (MRI) (e.g. \citealp{Flock2015,Flock2017,Riols2019}), zonal flows (e.g. \citealp{Uribe2015}), the compositional baroclinic instability \citep{Klahr2004} or gravitational instability \citep{Kratter2016} that could also account for the observations.  

\noindent One way to distinguish between the different scenarios and to understand possible planet-disc interactions is to directly image a young planet in its environment (e.g. \citealp{Keppler2018,Wagner2018}). Since this technique is only feasible for very few, massive objects, that are not affected by dust extinction \citep{Sanchis2020}, another promising method is to look for perturbations that are induced in the velocity field of the rotating gas. In this context, studying the gas component can help to access the different dynamical processes that are shaping the disc and reveal a number of previously undetected substructures. The density structure of dust grains that are typically probed by ALMA observations is determined by the gas dynamics. It is thus of paramount importance to directly access and characterize the gas kinematics to distinguish between various scenarios.

Different and complementary image analysis techniques to probe disc kinematics are being developed. For a geometrically thick disc around a single star that is both in radial and vertical hydrostatic equilibrium, the gas rotation velocity $v_{\mathrm{rot}}$ is given by 
\begin{equation}
\frac{v^2_{\mathrm{rot}}}{r} = \frac{GM_*r}{\left(r^2 + z^2\right)^{3/2}} + \frac{1}{\rho_{\mathrm{gas}}} \frac{\partial P}{\partial r}
\end{equation}
with $r$ being the cylindrical radius, $M_*$ the mass of the star, $\rho_{\mathrm{gas}}$ the gas density and $\partial P/\partial r$ representing the radial pressure gradient. Identifying deviations from Keplerian rotation can therefore be used to probe the local pressure gradient and to characterize the shape of the perturbation. Additional deviations may arise for a massive disc, due to its gravitational field. 

This technique has recently been used by \cite{Teaque2018,Teague2019} to constrain the gas surface density profile of the HD 163296 disc, leading to the kinematical detection of two embedded Jupiter-mass planets as well as significant meridional flows. In addition, \cite{Teague2018b} report a vertical dependence on the pressure maxima, studying the gas kinematics of AS 209. The deviations from Keplerian rotation are further used by \cite{Rosotti2020b} to measure the gas-dust coupling as well as the width of gas pressure bumps. \cite{Pinte2018,Pinte2019} detect \grq kink-' features in the iso-velocity contours of HD 163296 and HD 97048 data respectively, consistent with a Jupiter-mass planet ($\sim 2\,M_{\mathrm{Jup}}$). Tentative detections of such azimuthally located features have also been found in a few discs of the ALMA DSHARP large program \citep{Andrews2018,Pinte2020}, but more data are needed to confirm the robustness of such claim. Similarly, a possible signature for an embedded planet in the HD 100546 disc is presented by \cite{Casassus2019} who reveal a Doppler-flip in the residual kinematical structure after subtracting a Keplerian best-fit model, as expected from a planet-disc interaction model (e.g. \citealp{Perez2015,Perez2018}).

One type of discs, the so-called transition discs, are of particular interest, since they show evidence for dust (and gas) depleted inner regions (e.g. \citealp{Strom1989, Ercolano2017}). Sometimes treated as being in a transition phase from an optically thick disc to disc dispersal, transition discs may enable to probe various mechanisms that play a role during disc evolution and represent excellent candidates to catch planet formation in action. Detecting a planet in discs with cavities may link planet formation with fully formed planetary systems and put constraints on the formation processes and timescales.

In this work we study the transitional disc around CQ Tau. Following \cite{Gabellini2019}, who focused on the radial
profiles, characterization of the present dust and gas cavity and
possible formation mechanisms, we analyze the gas component of the CQ Tau disc both in terms of its velocity and temperature structure, finding significant spiral structures. The paper is organised as follows: In
\hyperref[sec:observations]{Sect.~\ref*{sec:observations}} we describe the observations and data reduction, whereas the observational results are presented in \hyperref[sec:results]{Sect.~\ref*{sec:results}}. A description and analysis of the spiral structure is shown in \hyperref[sec:analysis]{Sect.~\ref*{sec:analysis}} alongside with the method used to extract and model the gas kinematics. Our results are discussed in \hyperref[sec:Discussion]{Sect.~\ref*{sec:Discussion}} and summarised in \hyperref[sec:Summary]{Sect.~\ref*{sec:Summary}}. 
\section{Observations}\label{sec:observations}
\subsection{Target}
The variable star CQ Tau (UX Ori class) is a YSO of spectral type F2 located in the Taurus star forming region at a distance of $\sim 162\,\mathrm{pc}$ \citep{Gaia2018} $(\mathrm{RA} = 05^{\mathrm{h}}$:35$^{\mathrm{m}}:58.47^{\mathrm{s}}$, $\mathrm{Dec} = +24\degree:44':54.09'', \mathrm{J}2000)$. The intermediate mass star ($1.67\,\mathrm{M}_{\odot}$, \citealp{Lopez2006,Gabellini2019}) has an estimated age of $\sim 10\,\mathrm{Myr}$ and is surrounded by a massive circumstellar disc \citep{Natta2000}, which is found to have a high accretion rate of the order of $10^{-8} - 10^{-7}\,\mathrm{M}_{\odot}\,\mathrm{yr}^{-1}$ \citep{Donehew2011,Mendigut2012}.

The disc around CQ Tau represents one of the first discs whose mm continuum was observed with different instruments (e.g. OVRO interferometer \citep{Mannings1997}; PdBI \citep{Natta2000}; VLA \citep{Testi2001}) in order to constrain its dust properties. An analysis of the spectral slope at mm-wavelengths reveals that dust grains have grown to larger sizes than the typical ISM size \citep{Testi2001,Testi2003,Chapillon2008}. The average dust opacity coefficient is constrained by \cite{Banzatti2011} using VLA (1.3-3.6\,mm), PdBI (2.7-1.3\,mm) and SMA (0.87\,mm) observations, probing significant grain growth in the disc with up to cm-sized grains. \cite{Trotta2013} further find that larger grains are present in the inner disc with respect to the outer disc, indicating a variation of grain growth with radius. 

Subsequent high resolution gas and dust observations have revealed the CQ Tau disc to be a transition disc with an inner cavity. \cite{Tripathi2017} detect a gap in the 880$\,\mu \mathrm{m}$ continuum emission of new and archival SMA data and \cite{Pinilla2018} report a dust cavity of $\sim 46\,\mathrm{au}$ in ALMA observations of the mm continuum, fitting the intensity profile in the visibility plane. \cite{Gabellini2019} present recent ALMA observations, confirming a large cavity of 53\,au radius (peak of the Gaussian dust ring) in the 1.3$\,$mm continuum as well as a smaller gas cavity of 20\,au in the $^{13}$CO and C$^{18}$O emission, fitting the surface density profiles. The authors performed 3D hydrodynamical simulations which suggest a hidden planet of several $M_{\mathrm{J}}$ located at $\sim$ 20\,au as a possible cause for the observed gas and dust depleted regions. Even though such a planet could not be detected in combined Keck/NIRC2 and Subaru/AO188+HiCIAO observations of CQ Tau \citep{Uyama2019}, due to a lack of contrast (compare their Figure 3), the data reveal the presence of a small spiral seen in small dust grains on scales of 30-60\,au, that might be induced by a companion candidate.     
\subsection{Data reduction}\label{sec:almaobservations}
\begin{table*}[]
\centering
\caption{Characteristics of the data for the three lines $^{12}$CO, $^{13}$CO and C$^{18}$O.}\label{tab:dataInfo}
\begin{tabular}{l cccccc}
\hline
\hline
line & intrinsic resolution [kHz]& spectral resolution [$\mathrm{m}\,\mathrm{s}^{-1}$] & $\mathrm{b}_{\mathrm{maj}}$ [$''$]& $\mathrm{b}_{\mathrm{min}}$ [$''$] &
$\mathrm{b}_{\mathrm{PA}}$ [$\degree$] & rms channels [$\mathrm{Jy}\,\mathrm{beam}^{-1}$] \\
\hline
$^{12}$CO & 244 & 500 & 0.121 & 0.098 & 8.004 & 1.2e-3 \\
$^{13}$CO & 488 & 700 & 0.128 & 0.103 & 7.676 & 1.0e-3 \\
C$^{18}$O & 488 & 1000 & 0.129 & 0.103 & 8.814 & 0.7e-3 \\
\hline
\hline
\end{tabular}
\end{table*} 
We present 1.3\,mm ALMA observations of the CQ Tau system in band 6, combining datasets from cycle 2, 4 and 5 (2013.1.00498.S, PI: P\'erez; 2016.A.00026.S, 2017.1.01404.S., PI: Testi), previously presented at a lower angular resolution in \cite{Gabellini2019}, and detailed in \autoref{tab:obs_log}. 
These three projects have different antenna configuration, but they share a similar spectral setup. The longest baseline from the first project extends to 1091\,m, while for the latest two it is increased to 3700\,m and 8500\,m respectively, thus enhancing the spatial resolution. In all three observations the ALMA correlator was configured to observe the 1.3\,mm dust continuum emission, as well as the molecular lines $^{12}$CO $J = 2-1$, $^{13}$CO $J = 2-1$, and C$^{18}$O $J = 2-1$.

After applying ALMA standard pipeline calibration, we follow a similar processing as for the DSHARP data calibration \citep{Andrews2018}, using \texttt{CASA 5.4.1}. We start by flagging the channels located at $\pm 25\,\mathrm{km}\,\mathrm{s}^{-1}$ from each spectral line, and average the remaining channels to 125\,MHz width channels, which are combined with the data from the continuum spectral windows. As a next step, we align the dust continuum emission and check by comparison the flux calibration of each individual execution, to ensure they all have the same flux.

To enhance the signal-to-noise ratio, self-calibration is performed in two stages. First, we self-calibrate the shorter baseline dataset, corresponding to the Cycle 2 observations (dataset \#1 in \autoref{tab:obs_log}), by applying three steps of phase-only calibration using solution intervals of $300$, $120$ and $30\,$s, and one step of amplitude calibration using the whole observation time range as a solution interval. This self-calibrated short baseline dataset is then combined with the extended baseline datasets from the observations obtained in Cycles 4 and 5 (datasets \#2 and 3), and four phase calibrations with solution intervals of $900$, $360$, $150$, $90\,$s, as well as one amplitude calibration with solution interval of $360\,$s.

All the dust continuum emission calibration steps, including the centroid shifting and self-calibration tables, are then applied to the molecular line emission channels. The continuum emission is subtracted using the \texttt{uvcontsub} task, and image cubes are generated for each isotopologue (compare channel maps in \autoref{appendix:channels}) using a robust parameter of $0.6$ and Keplerian masking. This value was found to give the best trade off between spatial resolution and sensitivity. The Keplerian mask is calculated with the package \texttt{keplerian\_mask}\footnote{\url{https://github.com/richteague/keplerian_mask}}, using an inclination of 35\,\degree, position angle of 235\,\degree, distance of 162\,pc, stellar mass of 1.54\,$M_{\odot}$ and systemic velocity of $6.17\,\mathrm{km}\,\mathrm{s}^{-1}$. These values were chosen after some initial fits for the gas kinematics, explained later in \hyperref[sec:velocityRes]{Sect.~\ref*{sec:velocityRes}}. We further choose an inner and outer radius of 0 and 2$''$ respectively and convolve with the beam rescaled by 1.5 times its size. Some important characteristics of the data are given in \autoref{tab:dataInfo}.
\section{Observational results}\label{sec:results}
\subsection{Integrated and peak brightness temperature maps}\label{sec:maps}
\begin{figure*}[]
\centering
    \includegraphics[width=1.0\textwidth]{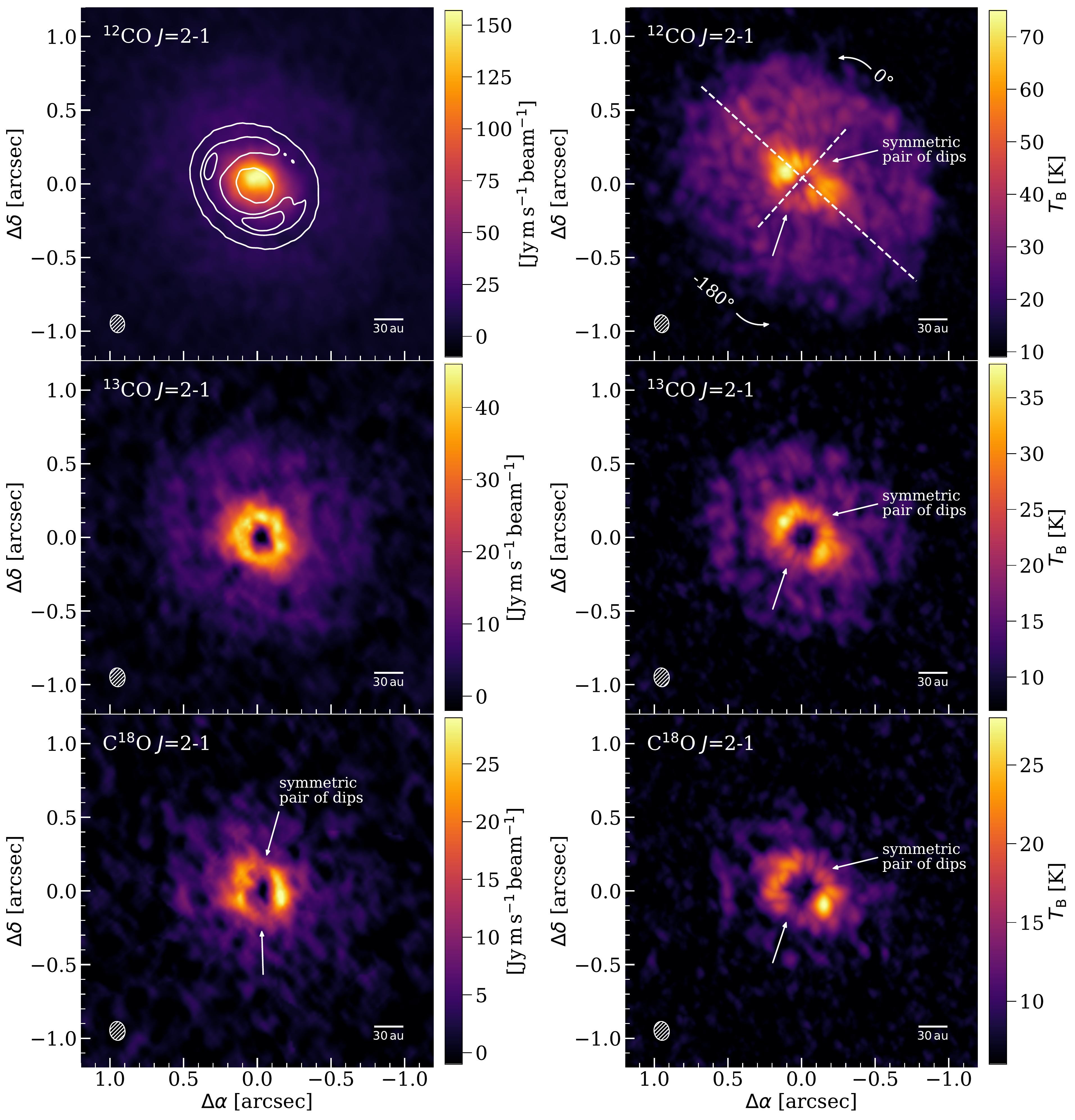}
    \caption{ALMA observations of the velocity integrated intensity (left panels) and peak brightness temperature (right panels) of the $^{12}$CO (top panels), $^{13}$CO (middle panels) and C$^{18}$O (bottom panels) $J=2-1$ transition. The conversion from flux to brightness temperature was performed with the Planck law. The contours of the continuum are overplotted on top of the $^{12}$CO integrated intensity at 20, 100 and 130\,$\sigma$ (1\,$\sigma$ = 11\,$\mu$Jy$\,$beam$^{-1}$). The synthesized beam is shown in the bottom left corner of each panel.}
    \label{fig:moments}
\end{figure*}
In \hyperref[fig:moments]{Fig.~\ref*{fig:moments}} we show the velocity integrated intensity maps (left panels) along with the peak brightness temperature maps (right panels) for the three different CO isotopologues. The underlying channel maps are presented in \autoref{appendix:channels}. The $^{12}$CO integrated intensity is overlaid by the contours of the 1.3\,mm continuum. To compute the integrated as well as the peak intensity maps we have used the \texttt{bettermoments} code described in \cite{bettermoments} and then converted from flux density units to units of Kelvin with the Planck law. In addition, the Keplerian mask described in \hyperref[sec:almaobservations]{Sect.~\ref*{sec:almaobservations}} is applied in the velocity integrated intensity maps to enhance the signal-to-noise. This results in a peak SNR of 40 for the $^{12}$CO, 20 for the $^{13}$CO and 14 for the C$^{18}$O velocity integrated intensity as well as 29 for the $^{12}$CO, 19 for the $^{13}$CO and 18 for the C$^{18}$O peak intensity.  

While the optically thick $^{12}$CO data do not trace any cavities in the gas distribution, a significant gas cavity is observed in the inner disc region as seen in the optically thinner $^{13}$CO and C$^{18}$O emission, shown also by \cite{Gabellini2019} at a lower angular resolution. In addition to \cite{Gabellini2019} we clearly note a drop of the peak intensity of all isotopologues and integrated intensity of C$^{18}$O by roughly 35-60\,\% (with respect to the peak) in the north-west and south-east part of the disc along the minor axis (symmetric pair of dips), that becomes especially prominent in the C$^{18}$O data and co-locates with a possible under-brightness in NIR scattered light (see Fig. 4 of \citealp{Uyama2019}). Similar to the NIR under-brightness, the drop of peak intensity appears to be more pronounced ($\sim$ 7-12\,\% deeper dip) in the south-east side of the disc (compare also \hyperref[sec:cuts]{Sect.~\ref*{sec:cuts}}). 

In the past such symmetric features have been linked to the presence of a misaligned inner disc, casting a shadow over the outer disc \citep{Marino2015,Facchini2018,Casassus2019}. However, the co-location of this under-brightness with the minor axis of the disc suggests caution, since beam dilution can lead to artificial azimuthal features due to the low compact emission of the line central channels. To test for this effect we have used the DALI (Dust And LInes \citealp{Bruderer2012,Bruderer2013}) model presented by \cite{Gabellini2019} and convolved the spectral image cubes with the beam of our observation. The resulting peak intensity map is shown in the left panel of \hyperref[fig:dilution]{Fig.~\ref*{fig:dilution}} of \autoref{appendix:dilution} for $^{12}$CO. Two clear dips appear along the minor axis, with the results being similar for the more optically thin lines. Thus, beam dilution can mostly account for the strong under-brightness seen in the data. Similar to the data, the upper side of the disc (including the north-west dip) appears slightly brighter in the model map. The disc vertical structure is likely playing a role here with the north-west side being the far side of the disk, thus associated to a larger projected emission area and a less severe beam dilution. In agreement with that, the upper side of the disc is also found to be brighter in some of the channels (compare \autoref{appendix:channels}).

The brightness temperature map of $^{12}$CO shows that the disk is also brighter, and thus likely warmer, in the north-eastern side. While $^{13}$CO does not show any strong east-west asymmetry, the disc is clearly brighter in the south-western side of C$^{18}$O, which instead of a higher temperature may trace a small over-density due to the lower optical depth of the line.  
\subsection{Radial and azimuthal cuts}\label{sec:cuts}
\begin{figure}
\centering
    \includegraphics[width=0.49\textwidth]{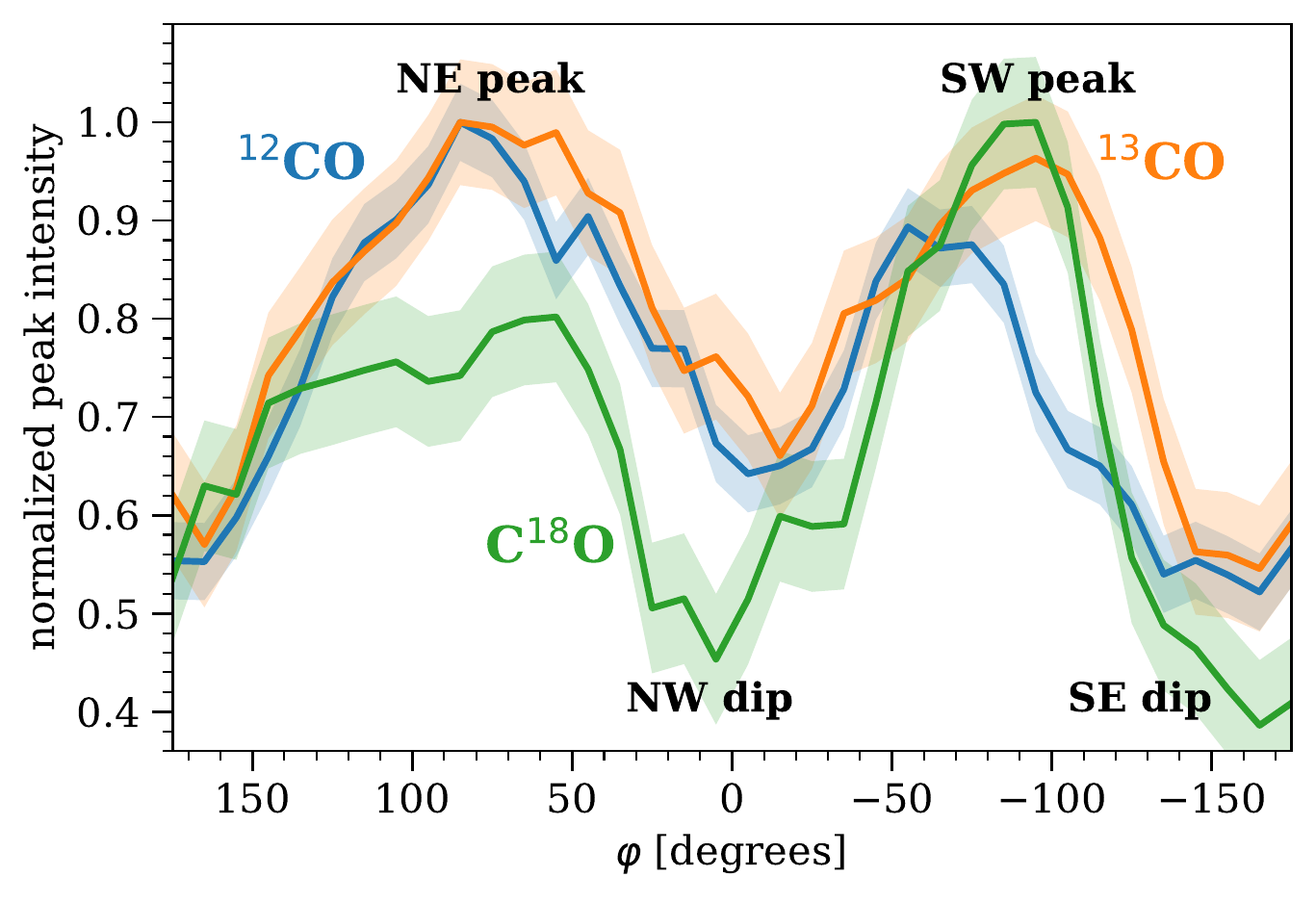}
    \caption{Azimuthal variation of the peak intensity for an annulus of 20-40\,au ($0.12-0.25''$), normalized to the peak value and shown for the three CO isotoplogues. The uncertainties are given as $1\,\sigma$.}
    \label{fig:polar}
\end{figure}
\hyperref[fig:polar]{Figure~\ref*{fig:polar}} presents the normalized azimuthal variations of the peak intensity (before $T_{\mathrm{B}}$ conversion) for the three CO isotopologues. Each profile is shown for the ring of strongest intensity between $0.12-0.25''$ ($\sim $ 20-40\,au) with the uncertainty being the rms computed on the peak intensity map. To rotate and deproject the maps an inclination and position angle of $35\,\degree$ and $235\,\degree$ respectively (compare results of \hyperref[sec:ThinDiscChapter]{Sect.~\ref*{sec:ThinDiscChapter}}) were used. As discussed above we indeed find an opposite east-west asymmetry in the curves for $^{12}$CO and C$^{18}$O, while the intensity of $^{13}$CO is relatively symmetric along the azimuth. The intensity of the left (east) peak compared to the right (west) peak is roughly roughly 10\,\% higher for the $^{12}$CO and 25\,\% lower for the C$^{18}$O peak intensity. This matches the brightness and temperature differences seen in the maps of \hyperref[fig:moments]{Fig.~\ref*{fig:moments}}, where the east side of the disc is brighter in $^{12}$CO but fainter in C$^{18}$O. 

\noindent In addition, the profiles underline that the under-brightness appears stronger in the south-east of the disc. For comparison, the azimuthal profiles derived from the DALI model are shown in the right panel of \hyperref[fig:dilution]{Fig.~\ref*{fig:dilution}} for all three isotopologues. In contrast to the data no strong east-west asymmetry is seen in the peaks. The peak intensity further drops by 35-45\,\% at the dips with the south-east dip being about 2-6\,\% deeper compared to the north-west dip. The asymmetry found in the under-brightness is thus more pronounced in the data. Together with the under-brightness seen in NIR, where beam dilution cannot be invoked to explain the latter, this supports the assumption that an additional shadowing may occur in the south side of the disc. While the continuum shows two clumps (top left panel of \hyperref[fig:moments]{Fig.~\ref*{fig:moments}}) they are present at a different location than the brightness asymmetries as well as the symmetric pair of dips and are thus unlikely to account for the variations.

\hyperref[fig:intensity]{Figure~\ref*{fig:intensity}} displays the normalized radial intensity profiles for the three CO isotopologues as well as the 1.3\,mm continuum. The curves are obtained from the azimuthally averaged intensity per annuli of size $0.02'' (\sim 3.2\,\mathrm{au})$ from the peak intensity maps (before $T_{\mathrm{B}}$ conversion), again using an inclination and position angle of $35\,\degree$ and $235\,\degree$ respectively. The error bars are calculated as the standard deviation per annulus divided by the square root of number of independent beams in the annulus. Compared with the profiles shown by \cite{Gabellini2019} for the integrated intensity, we notice a drop of the peak intensity in the radial profile of C$^{18}$O between $\sim$ 65-85\,au of about 3\,\% in addition to the intensity drop at the $\sim 20\,\mathrm{au}$ cavity in $^{13}$CO and C$^{18}$O. A corresponding slight dip is present in the $^{13}$CO peak intensity profile. Being more optically thin, C$^{18}$O is mostly tracing the column density. Therefore the observed feature may be indicative of a depleted region around 75\,au, possibly carved by an unseen companion. Continuum absorption is unlikely to account for the dip since the peak of the continuum flux lies around 52\,au rather than 75\,au. Another explanation may be the enhancement of emission around $\sim$ 90\,au at the edge of the continuum, rather than a dip, potentially caused by an enhanced desorption of CO ices by increased UV or a temperature inversion (e.g. \citealp{Cleeves2016,Facchini2017}). 
\section{Analysis}\label{sec:analysis}
\subsection{Temperature structure}\label{sec:tempres}
\begin{figure}
\centering
    \includegraphics[width=0.49\textwidth]{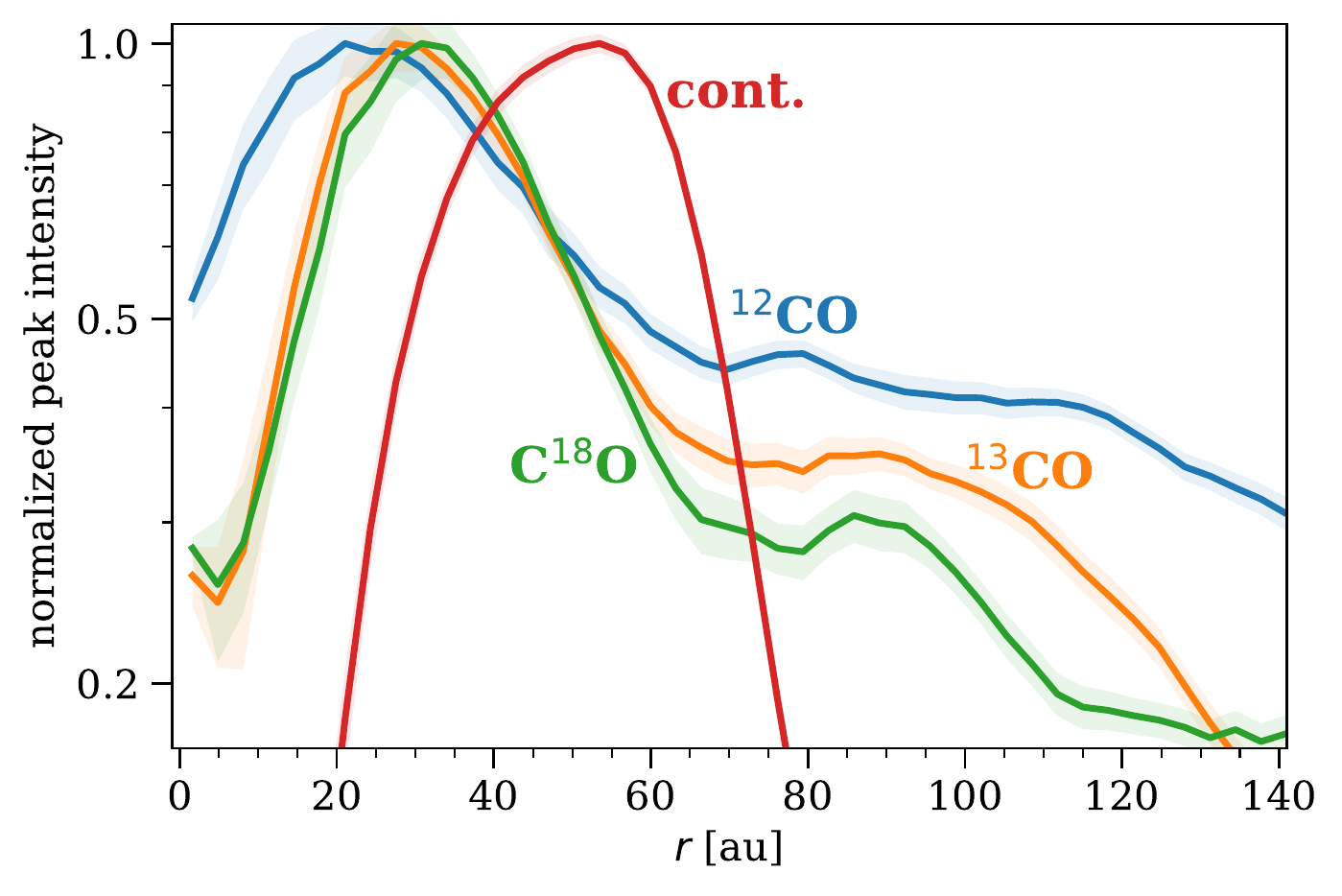}
    \caption{Azimuthally averaged, radial intensity profiles of the continuum (red), $^{12}$CO (blue), $^{13}$CO (orange) and C$^{18}$O (green) data. The profiles are derived from the corresponding peak intensity maps and normalized to the peak value.}
    \label{fig:intensity}
\end{figure}
\begin{figure}
\centering
    \includegraphics[width=0.49\textwidth]{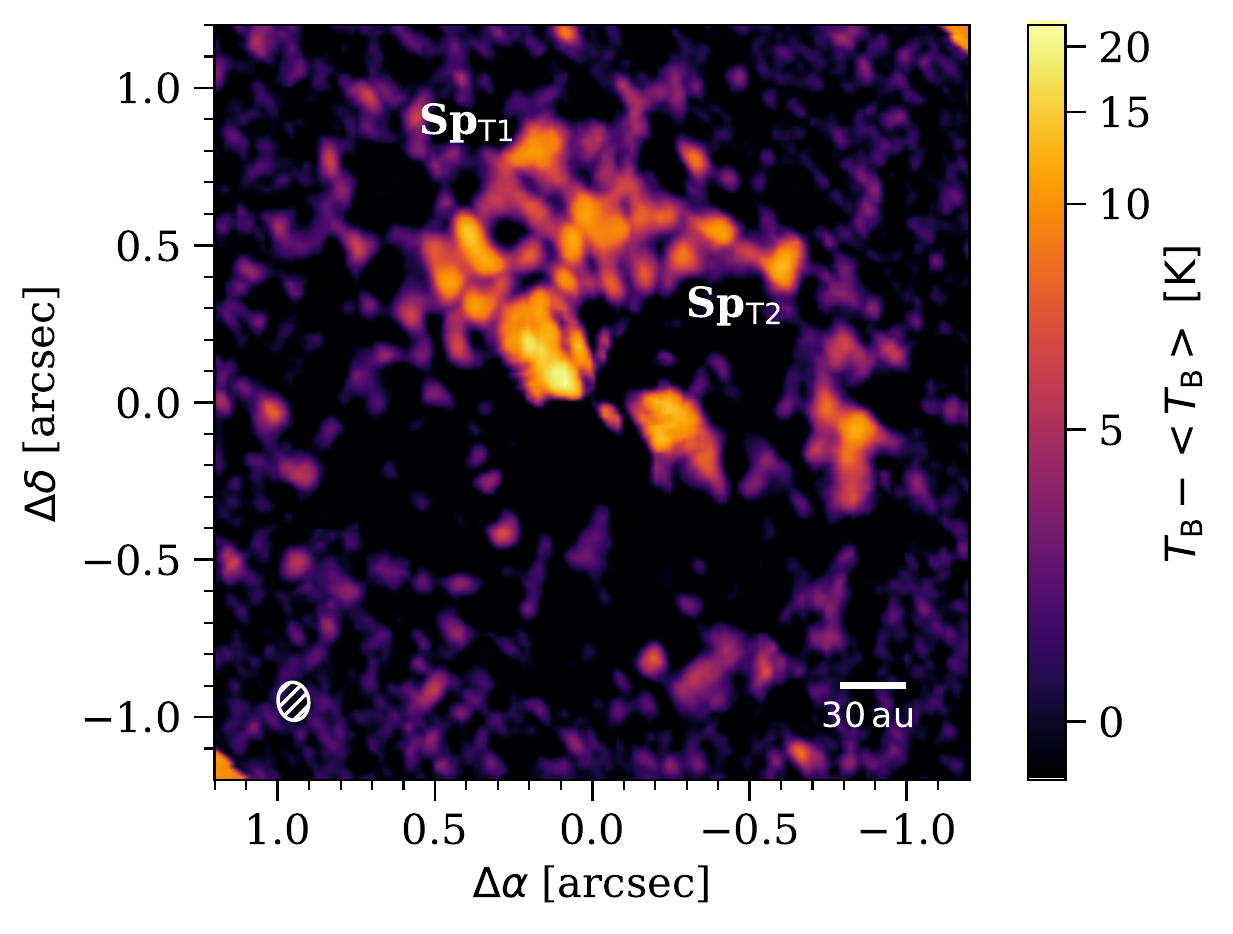}
    \caption{Residuals of the $^{12}$CO brightness temperature after subtraction of an azimuthally averaged radial profile. Two spirals SP$_{\mathrm{T1}}$ and SP$_{\mathrm{T2}}$ are spanning an azimuth of $\sim $ 100\,\degree and > 180\,\degree respectively between $\sim$ 10-180\,au.}
    \label{fig:residualsT}
\end{figure}

Since the $^{12}$CO emission is optically thick and in LTE at these low rotational transitions, the brightness temperature (top right panel of \hyperref[fig:moments]{Fig.~\ref*{fig:moments}}) can be used as a probe of the gas kinetic temperature. In this context the gas temperatures up to 75\,K that we observe are as they would be expected in the upper disc layers \citep{Bruderer2014}. To uncover small perturbations in this temperature structure, similar to \cite{Teague2019Spiral} we subtract an azimuthally averaged radial $T_{\mathrm{B}}$ profile similar to the one shown in the bottom panel of \hyperref[fig:intensity]{Fig.~\ref*{fig:intensity}}. This leaves significant spiral structure in the resulting residuals as shown in \hyperref[fig:residualsT]{Fig.~\ref*{fig:residualsT}}. Two clear spirals are observed, a smaller one (Sp$_{\mathrm{T1}}$) spanning an azimuth of $\sim 100\,\degree$ between $\sim 10-180\,\mathrm{au}$ and a larger spiral (Sp$_{\mathrm{T2}}$) covering more than half an azimuth at a similar radial extent. Both spirals have the same orientation. The small spiral seen in the NIR by \cite{Uyama2019} at radii of $\sim$ 30-60\,au co-locates with the anchoring point of the large spiral observed here (compare \hyperref[fig:SpiralT]{Fig.~\ref*{fig:SpiralT}} in \hyperref[sec:Spiral]{Sect.~\ref*{sec:Spiral}}).
\begin{figure*}
\centering
    \includegraphics[width=\textwidth]{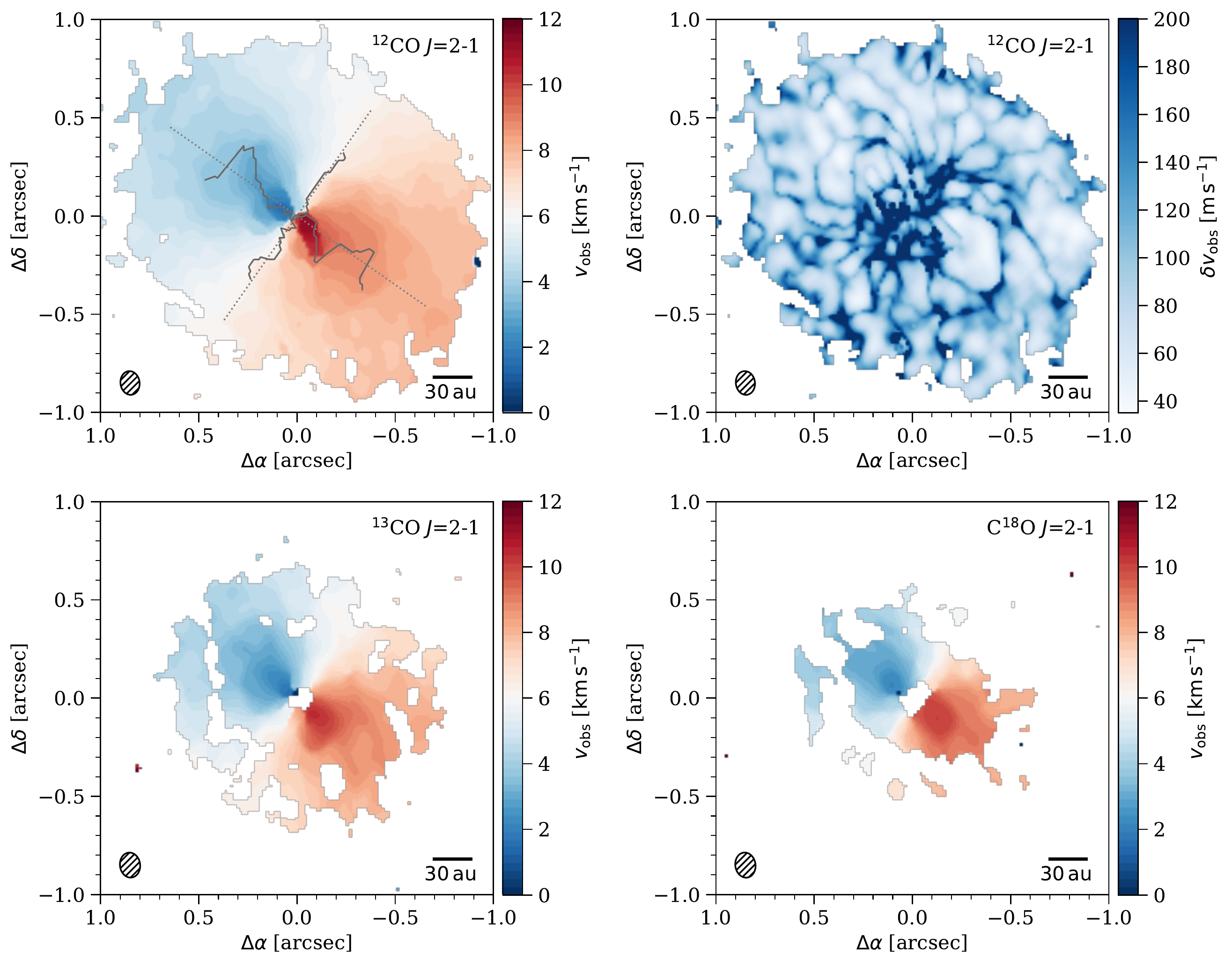}
    \caption{Rotation map of the three CO isotopologues and the corresponding uncertainties of $^{12}$CO calculated with \texttt{bettermoments} (top right panel). Regions below 4\,$\sigma$ ($^{12}$CO) and 3.5\,$\sigma$ ($^{13}$CO, C$^{18}$O) are masked out. The maximum and minimum velocities along the red- and blue-shifted minor and major axis are overlaid with grey lines over the $^{12}$CO rotation pattern.}
    \label{fig:velocitymaps}
\end{figure*}
\subsection{Velocity structure}\label{sec:velocityRes}
\begin{figure*}
\centering
    \includegraphics[width=\textwidth]{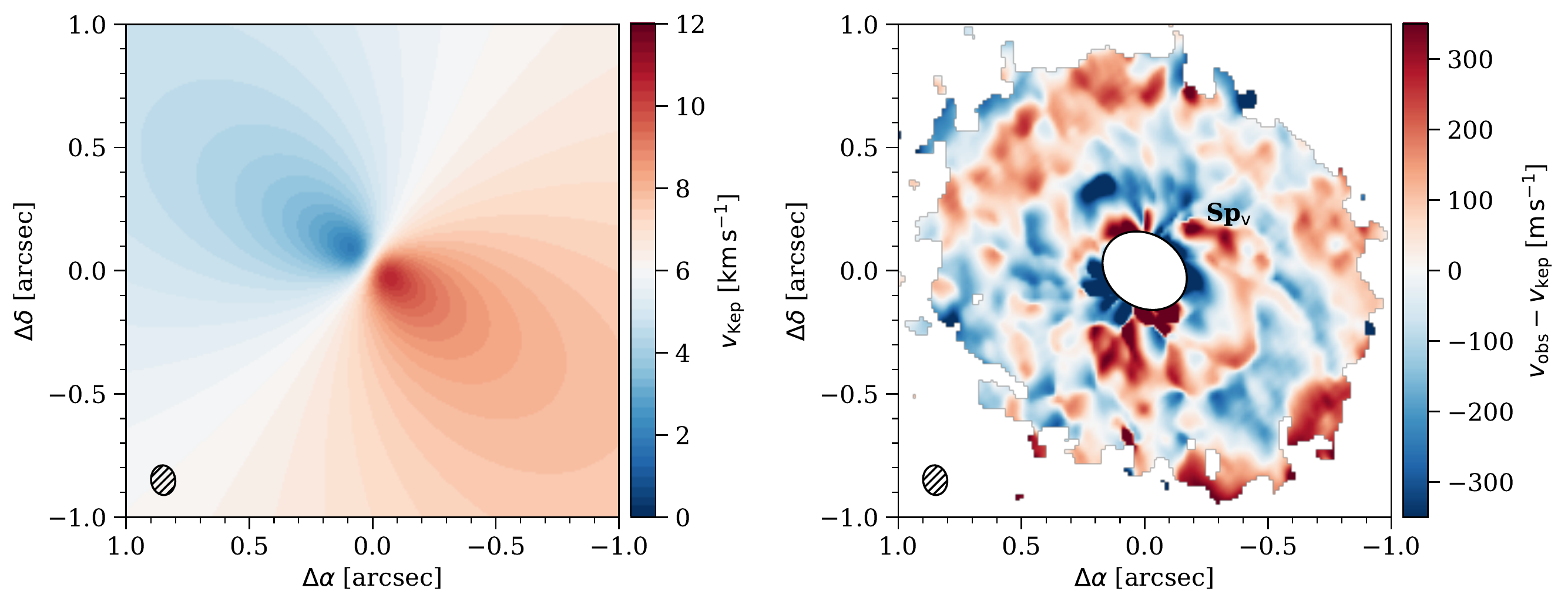}
    \caption{Best-fit Keplerian rotation model (left panel) and the residuals calculated by subtracting the model from the observed rotation velocity (right panel), shown for $^{12}$CO. The residuals inside of 1.5 the FWHM of the beam are masked out.}\label{fig:ResidualsRun1}
\end{figure*}
The \texttt{bettermoments} package can also be used to generate the observed rotation velocity of the gas (similar to a moment 1 map). The code fits a quadratic model to the brightest pixel as well as the two neighbouring pixels to find the centroid of the line in pixel coordinates. Compared to other methods this approach is more robust to noise or errors in the line shape, allowing a precision that is greater than the velocity resolution. The resulting velocity structure of the disc, masking regions below 4\,$\sigma$ for $^{12}$CO and 3.5\,$\sigma$ for $^{13}$CO and C$^{18}$O, is shown in \hyperref[fig:velocitymaps]{Fig.~\ref*{fig:velocitymaps}}. Besides the velocity field of all three isotopologues the corresponding error map of $^{12}$CO is included in the top right panel. These statistical uncertainties are calculated by linearizing and propagating the uncertainty from the fluxes to the centroid estimate. For most regions the achieved precision is well below the channel width of $500\,\mathrm{m}\,\mathrm{s}^{-1}$. In the central regions the uncertainties increase due to beam smearing. 

The isovelocities of the outer disc in the $^{12}$CO velocity map match those of a Keplerian flat disc model (e.g. \citealp{Rosenfeld2013}), whereas significant distortions can be noticed in the inner disc (up to $\sim$ 0.5$''$) with the kinematics in the centre being slightly twisted and the blue- and red-shifted parts bending in opposite directions. In the top left panel of \hyperref[fig:velocitymaps]{Fig.~\ref*{fig:velocitymaps}} the maximum and minimum velocities along the red- and blue-shifted minor and major axis are overlaid, emphasizing the non-Keplerian term present in the inner disc. For the case of a razor-thin Keplerian disc a dipole morphology would be expected that is symmetric about the semi-major axis. The isovelocities further hint towards a rather non- or slightly elevated or flared emission surface since the lobes of the rotation pattern are overall not distinctively bent away (in one direction) from the disc major axis, although this may be resulting from the perturbing spiral structure. We still attempted to fit for the emission surface, however none of the fits converged. In the following we will thus focus on a razor-thin disc Keplerian model. 

The rotation pattern of $^{13}$CO shows a similar twisting and bending as the $^{12}$CO emission while no substructure can be discerned in the less bright C$^{18}$O.    
\subsubsection{Analysis of the gas rotation velocity}\label{sec:AnaVel}
To analyze the gas kinematics of the disc around CQ Tau and characterize the apparent deviations from Keplerian velocity we fit a Keplerian profile
\begin{equation}\label{eq:kepler}
v_{\mathrm{rot}} (r, \phi) = \sqrt{\frac{G M_*}{r}}  \cdot \cos{\phi} \cdot \sin{i} + v_{\mathrm{LSR}} 
\end{equation}
with $(r,\phi)$ being the deprojected cylindical coordinates, $i$ the inclination of the disc and  $v_{\mathrm{LSR}}$ the systemic velocity to the rotation map of $^{12}$CO shown in \hyperref[fig:velocitymaps]{Fig.~\ref*{fig:velocitymaps}} using the \texttt{eddy} code \citep{eddy}. The associated uncertainties are included in the fit. In order to deproject the sky-plane coordinates $(x,y)$ into the midplane cylindrical coordinates $(r,\phi)$, the disc centre $(x_0,y_0)$, $i$ and the disc position angle PA are used. The latter is measured between the north and the red-shifted semi-major axis in an easterly direction. As a first step, the starting positions of the free fit parameters are optimized with \texttt{scipy.optimize} with their posterior distributions estimated using the MCMC sampler. In this context we used 200 walkers, 5000 steps to burn in and 5000 additional steps to sample the posterior distribution function. For all of our models we assumed flat priors that were allowed to vary over a wide range. The uncertainties of the posterior distributions represent the 16th to 84th percentiles about the median value.

In addition to the razor-thin disc model, a parameterization for the emission surface as well as a warped structure can be included in the model.   
The results of our modelling are reported in the following and summarized in \autoref{tab:ParametersThin} of \autoref{appendix:table}.
\subsubsection{Razor-thin disc model}\label{sec:ThinDiscChapter}
\begin{figure*}
\centering
\includegraphics[width=1.0\textwidth]{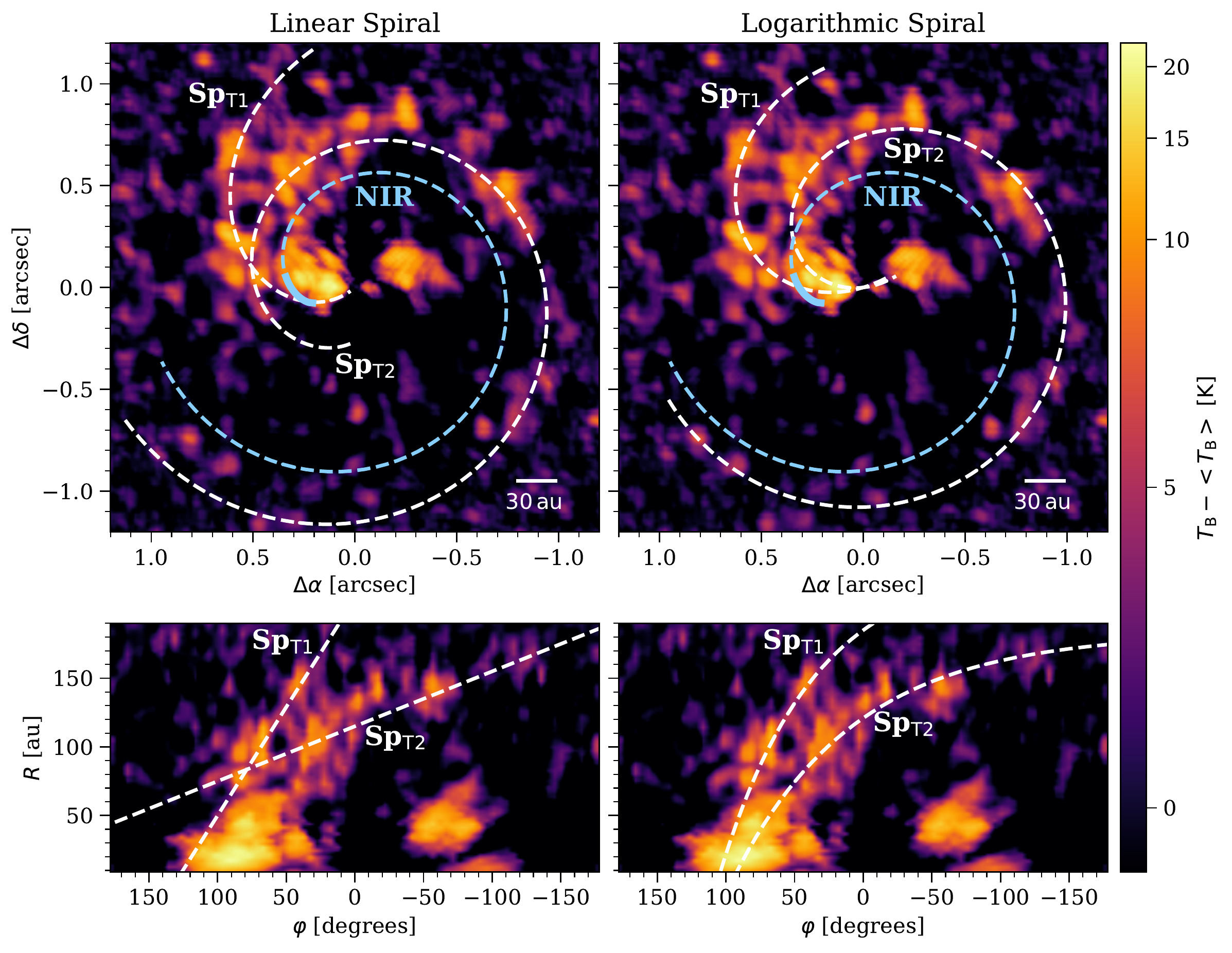}
\caption{Deprojected and rotated (top panels) as well as polar-deprojected (bottom panels) $T_{\mathrm{B}}$ residuals of $^{12}$CO with overlaid Archimedean and logarithmic spirals. The blue dashed lines show the fit of the spiral observed in the NIR by \citet{Uyama2019} extrapolated to large azimuthal angles. The solid blue line in inner region of the spiral highlights the region used by \citet{Uyama2019} to obtain the fit.}\label{fig:SpiralT}
\end{figure*}
\begin{figure*}
\centering
\includegraphics[width=1.0\textwidth]{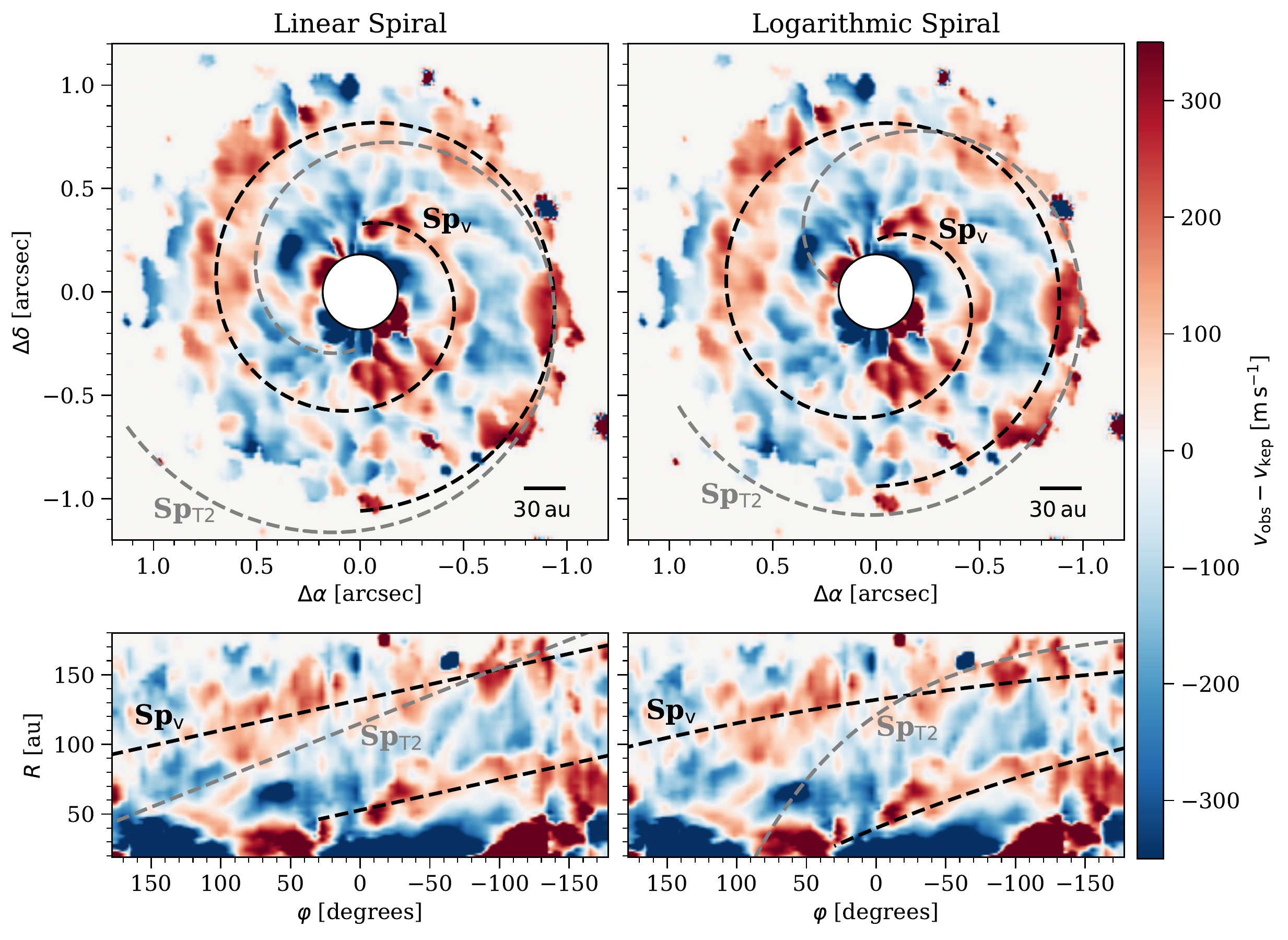}
\caption{Deprojected and rotated (top panels) as well as polar-deprojected (bottom panels) velocity residuals of $^{12}$CO with overlaid Archimedean and logarithmic spirals. The spiral Sp$_\mathrm{T2}$ observed in $T_{\mathrm{B}}$ is overlaid as a grey dashed line. The residuals inside of 1.5 the FWHM of the beam are masked out.}\label{fig:Spiralv}
\end{figure*}
For the razor-thin disc models we fixed the object's distance and inclination to 162\,pc and 35\,$\degree$ (\citealp{Gabellini2019}) respectively, and fitted for the disc centre $(x_0, y_0 \in \{-0.5'',0.5''\})$, systemic velocity $(v_{\mathrm{LSR}} \in \{-5\,\mathrm{km}\,\mathrm{s}^{-1},20\,\mathrm{km}\,\mathrm{s}^{-1}\})$, stellar mass $(M_* \in \{0.1\,M_{\odot},5\,M_{\odot}\})$ and disc position angle (PA $\in \{-360\,\degree,360\,\degree$\}).  

In the first two runs we attempted to fit the entire disc, choosing an outer radius of 1.0$''$ (162\,au) to exclude possible noise at the disc's edge. For the second run we further set an inner boundary of 0.25$''$ (40\,au), which corresponds to the inner edge of the Gaussian ring of the dust continuum, as obtained by \cite{Gabellini2019}. Both set-ups result in very similar fit parameters, yet returning a slightly larger stellar mass when the disc centre is excluded. In addition, we tried to fit specific regions of the disc, including only the inner disc (run 3,4), outer disc (run 5,6) as well as annuli of size 0.2$''$ (run 7-10). Overall we find that $v_{\mathrm{LSR}}$ is slightly increasing towards the outer disc, while PA is relatively constant. The largest scatter is found for the stellar mass $M_*$, that is ranging from 1.47 to 1.65\,M$_{\odot}$, driven by the model trying to account for the non-Keplerian structure in the rotation map. 

All thin disc models rapidly converged with a Gaussian posterior distribution function (PDF), resulting in relatively similar residuals when the model is subtracted from the velocity data. We tried both convolving the models with the beam of the observation and not using the convolution, with both approaches returning comparable results. In this context we note that convolving channel maps prior to collapsing them into the rotation map would be the better approach than convolving the model map as it is done in \texttt{eddy}. Generating channels maps, as opposed to a simple rotation map, however requires far more model assumptions which is why we choose not to do it. The effects are negligible outside the disc centre (i.e. outside $\sim$2 beam FWHM) and thus do not significantly affect our results.

The posterior distributions presented in \autoref{tab:ParametersThin} show very small and likely underestimated uncertainties, especially in context of the scatter that is found in the stellar mass. One possibility is that the uncertainties in the velocity centroid are underestimated. We thus performed an additional run of model 2 with the velocity errors increased by a factor of 10. This returns very similar fit parameters with the uncertainties also being a factor 10 larger, however still too small to account for the observed scatter. Therefore the small uncertainties cannot only be explained by underestimated velocity errors but result from systematic uncertainties in our model.

In \hyperref[fig:ResidualsRun1]{Fig.~\ref*{fig:ResidualsRun1}} the results from run 2 are presented, including the best-fit model (left panel) and the corresponding residuals (right panel) after subtraction from the data. The model corresponds to a position angle PA $ = 235\,\degree$, systemic velocity $v_{\mathrm{LSR}} = 6173\,\mathrm{m}\,\mathrm{s}^{-1}$ and a stellar mass of $M_* = 1.57\,$M$_{\odot}$. While the residuals are less than about 5\,\% of the velocity data outside of $\sim 0.2''$, they (partly) grow to more than 50\,\% in the very inner disc ($< 0.1''$) which suffers strongly of beam smearing effects \citep{Teague2018_2,Teague2016}. We thus masked out the residuals inside of 1.5 the FWHM of the beam. They clearly reveal the non-Keplerian structure of the rotation velocity, showing significant spiral features that cover more than one azimuth at radii of 40-180\,au, with the same orientation and a similar location as the spirals observed in the gas temperature. 

Since twisted kinematics are sometimes linked to the presence of a misaligned inner disc, we performed several runs, adding the parameterization of a potential warp in the (flat disc) model. Yet none of these models converged. Besides being limited by the spatial resolution of the data ($0.121'' \times 0.098''$) the kinematics are strongly dominated by the large spiral structure in the outer disc. Thus a small feature such as a warp in the very inner disc regions can not be fit by our simple model. For the same reason we were not able to obtain any constraints on the emission surface.
\subsection{Analysis of the spiral structure}\label{sec:Spiral}
\begin{figure*}
\centering
\includegraphics[width=1.0\textwidth]{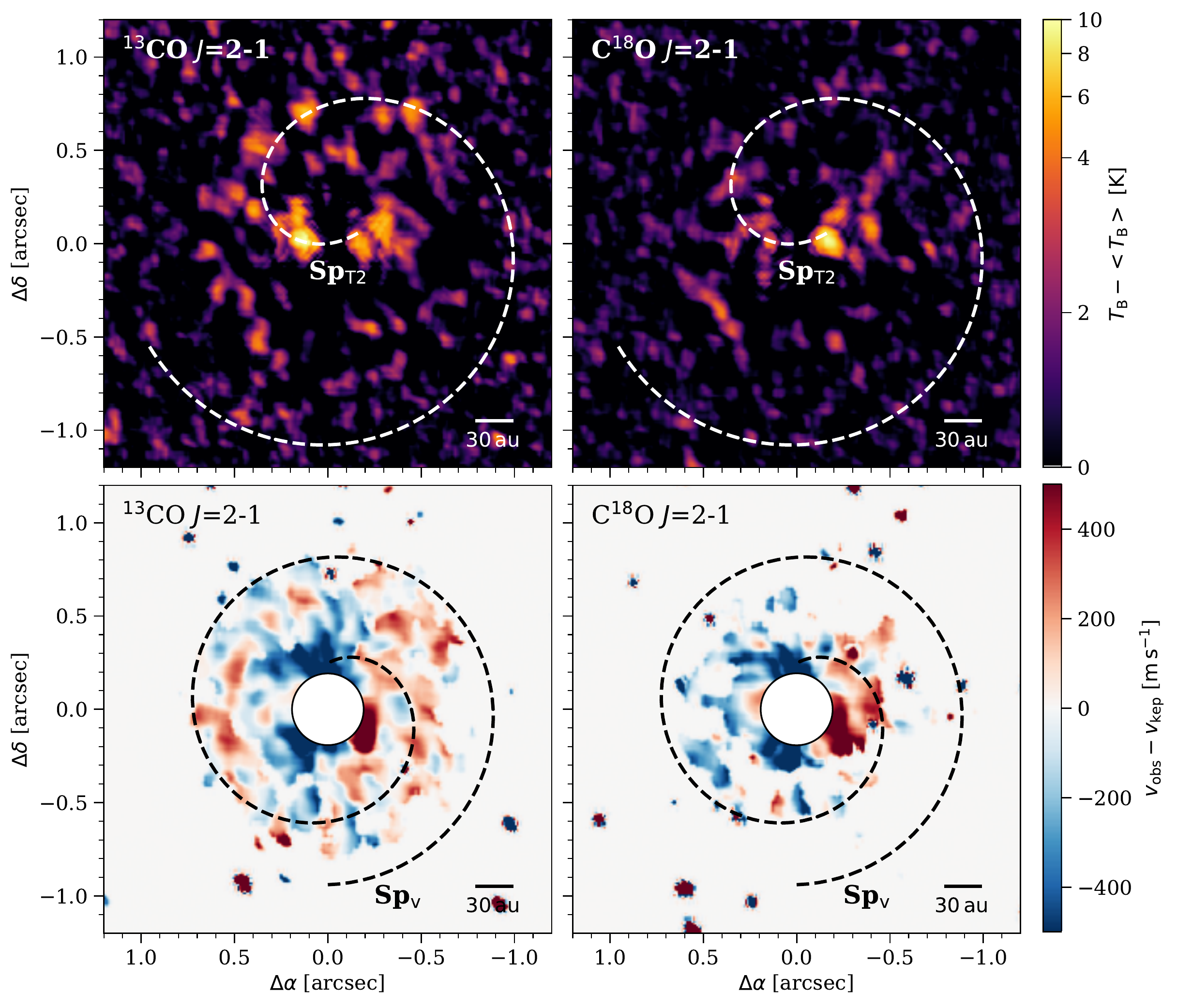}
\caption{Deprojected and rotated $T_{\mathrm{B}}$ (top) and velocity (bottom) residuals of $^{13}$CO (left) and C$^{18}$O (right). The logarithmic spirals Sp$_\mathrm{T2}$ and Sp$_\mathrm{v}$ observed in $^{12}$CO are overlaid. The residuals inside of 1.5 the FWHM of the beam are masked out in the velocity.}\label{fig:SpiralTothers}
\end{figure*}
\begin{figure*}
\centering
\includegraphics[width=1.0\textwidth]{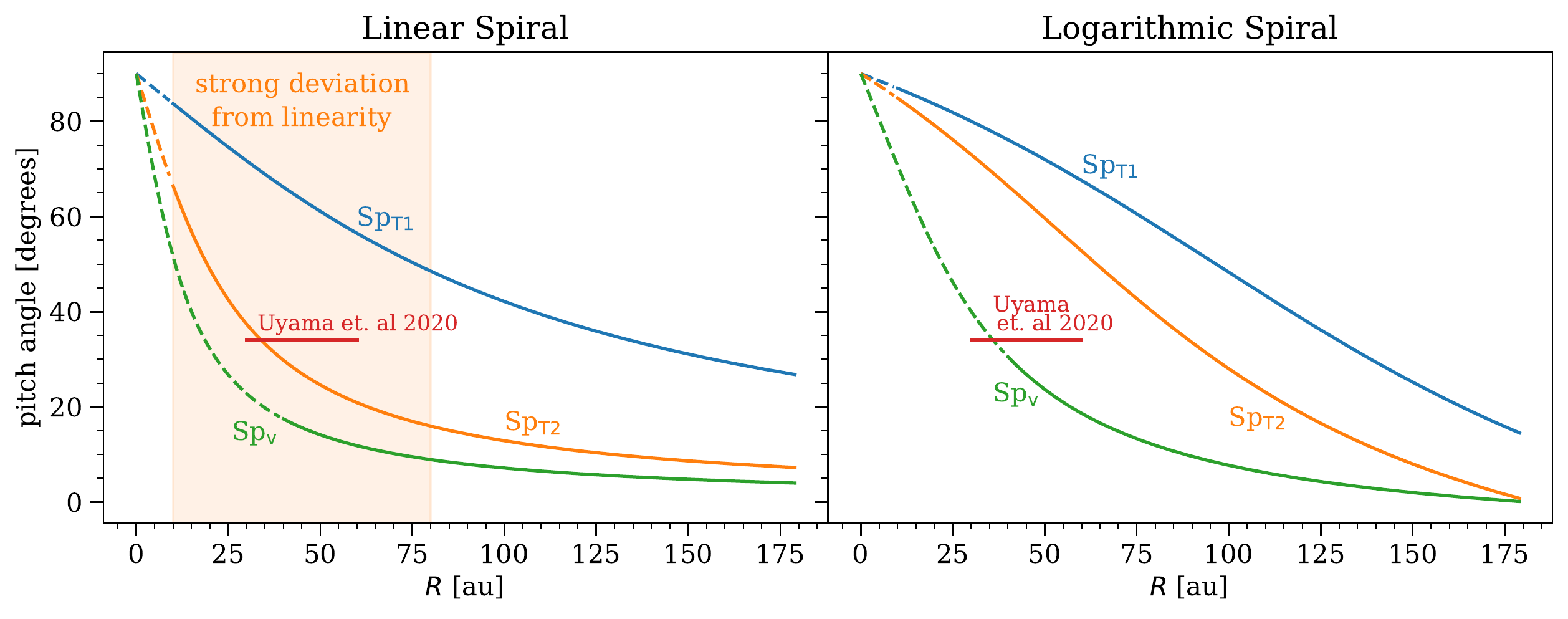}
\caption{Pitch angle of the three spirals observed in the $T_{\mathrm{B}}$ and rotation velocity residuals of $^{12}$CO, shown for the linear and logarithmic spiral. The constant pitch angle of the NIR spiral found by \cite{Uyama2019} is included as a reference. The shaded region represents the disc regions where the linear fit fails to reproduce the morphology of spiral Sp$_{\mathrm{T2}}$.}\label{fig:pitchangles}
\end{figure*}
Three significant spirals are observed in the residuals presented in \hyperref[sec:tempres]{Sect.~\ref*{sec:tempres}} and \hyperref[sec:velocityRes]{Sect.~\ref*{sec:velocityRes}}, two in the brightness temperature and one in the rotation velocity of $^{12}$CO. All spirals show the same orientation, suggesting a counter-clockwise rotation of the disc if the spirals are trailing, and they cover a large azimuth and radial extent. A counter-clockwise rotation implies that the south-east side of the disc is closer to the observer, which should be visible in the rotation map as a bending of the high velocity components towards the north. While this is true for the blue-shifted side, the spiral may be driving such a large velocity perturbation, that the red-shifted side is bending the other way, then resembling the presence of a warp. The geometry of the disc agrees with the SE dip in the peak brightness temperature map being dimmer due to beam dilution (see \hyperref[sec:maps]{Sect.~\ref*{sec:maps}}). 

\begin{table}
\caption{Parameters for the by-eye parameterization with a linear and logarithmic spiral.}\label{tab:spirals}
\begin{tabular}{l ccccc}
\hline
\hline
\multicolumn{1}{c}{\textit{\textbf{}}} & \multicolumn{2}{c}{\textbf{Linear Spiral}} & \multicolumn{3}{c}{\textbf{Logarithmic Spiral}} \\
\hline
\vspace{0.1cm}
Spiral & $a$ [$''$] & $b$ [$''$/rad] & $a$ [$''$] & $b$ [$''$] & $k$ [1/rad] \\
\hline
\multicolumn{5}{l}{\textbf{brightness temperature}} \\
Sp$_\mathrm{T1}$ & 1.284 & -0.559  & 1.451 & -0.309 & 0.831\\
Sp$_\mathrm{T2}$ & 0.710 & -0.142 & 1.130 & -0.383 & 0.642\\
\\
\multicolumn{5}{l}{\textbf{rotation velocity}} \\
\vspace{0.1cm}
Sp$_\mathrm{v}$ & 0.815 & -0.078 & 1.123 & -0.309 & 0.166\\
\hline
\hline
\end{tabular}
\end{table} 
We reproduce the spiral morphology with different functional forms, in particular an Archimedean (linear) spiral  
\begin{equation}
r = a + b\phi 
\end{equation}
as well as a spiral
\begin{equation}\label{eq:logSp}
r = a + b\mathrm{e}^{k\phi}
\end{equation}
where $r$ represents the radius and $\phi$ the polar angle of the spiral. Since \hyperref[eq:logSp]{Eq.~\ref*{eq:logSp}} is similar to the equation of a logarithmic spiral, we will refer to it as such in the following. The resulting parameters are presented in \autoref{tab:spirals} and the corresponding spirals are shown in \hyperref[fig:SpiralT]{Fig.~\ref*{fig:SpiralT}} for the brightness temperature and in \hyperref[fig:Spiralv]{Fig.~\ref*{fig:Spiralv}} for the rotation velocity. In both cases we plot the deprojected and rotated maps, using the inclination and position angle found in \hyperref[sec:ThinDiscChapter]{Sect.~\ref*{sec:ThinDiscChapter}} (top plots). Again the inner disc regions are masked out inside of 1.5 the FWHM of the beam for the velocity residuals. Additionally the polar-deprojected maps are shown (bottom plots). The spiral fits are overlaid as Sp$_\mathrm{T1}$ and Sp$_\mathrm{T2}$ in the temperature and as Sp$_\mathrm{v}$ in the velocity with white and black dashed lines. 

Both functions provide a good and very similar parameterization for the small temperature spiral Sp$_{\mathrm{T1}}$, with the overall linear nature becoming clear when looking at the polar-deprojected map. For the second, large temperature spiral Sp$_{\mathrm{T2}}$, the linear spiral is not able to account for the curvature obvious in the polar-deprojected plot and is thus missing the inner part of the disc. The logarithmic parameterization on the other hand is able to better represent the anchoring point of the spiral. The velocity spiral Sp$_{\mathrm{v}}$ is again well represented by both functions. No large differences can be seen between the two parameterizations, with the overall nature being very close to linear. As a comparison the Sp$_\mathrm{T2}$ spiral is also included in the velocity residuals. While the two spirals are co-located in the outer regions, they deviate from each other towards the inner disc for both spiral functions. 

The spiral found in the NIR is overlaid on the brightness temperature residuals in \hyperref[fig:SpiralT]{Fig.~\ref*{fig:SpiralT}}. Here the solid line represents the location of the NIR spiral, while the dashed line shows the extrapolation of the fit performed by \cite{Uyama2019} for this spiral with a function $r = a + b\phi^n$. The NIR spiral matches the anchoring point of the temperature spirals and follows the course of Sp$_\mathrm{T2}$ at the inner edge, suggesting them to be connected.  

In addition to the $^{12}$CO data we search for features in the residuals of $^{13}$CO and C$^{18}$O, presented in \hyperref[fig:SpiralTothers]{Fig.~\ref*{fig:SpiralTothers}} for $T_{\mathrm{B}}$ (top panels) and the velocity (bottom panels). The residuals are calculated again by subtracting an azimuthally symmetric model, using the best-fit model from run 2 for the velocity. The logarithmic spirals Sp$_\mathrm{T2}$ and Sp$_\mathrm{v}$ observed in $^{12}$CO are overlaid for comparison. Even though no clear spiral structure can be found in the brightness temperature of either $^{13}$CO or C$^{18}$O, the launching point of the spiral is clearly visible in the $^{13}$CO residuals and some indication of a spiral matching the logarithmic parameterization of Sp$_\mathrm{T2}$ is present. Similarly, the velocity residuals of $^{13}$CO also show indication for a spiral similar to that observed in $^{12}$CO, yet slightly more tightly wound, while no substructure can be distinguished in C$^{18}$O.     

Using
\begin{equation}
\tan \beta = \Bigg|\frac{dr}{d\phi}\Bigg| \cdot \frac{1}{r} 
\end{equation}
we calculate the pitch angle $\beta$ for all three spirals. The resulting angles are shown in \hyperref[fig:pitchangles]{Fig.~\ref*{fig:pitchangles}} for the linear and logarithmic fit. Here the solid lines are shown at the location where the spirals are present, while the dashed lines represent the extrapolation towards smaller radii. We further included the constant pitch angle found by \cite{Uyama2019}, who fitted a logarithmic spiral of $r = b\mathrm{e}^{k\phi}$ to a spiral feature observed in the NIR. As mentioned above this spiral is located at the anchoring point of spiral Sp$_\mathrm{T2}$, yet spanning a smaller radial extent of $\sim$ 30-60\,au and azimuth of roughly 50\,$\degree$. The pitch angle of the NIR spiral, which is given as $\sim 34\,\degree$, seems consistent with the pitch angle found from the linear parameterization for Spiral Sp$_\mathrm{T2}$, however in these radial regions the linear fit failed to reproduce the spiral morphology (shaded region in \hyperref[fig:pitchangles]{Fig.~\ref*{fig:pitchangles}}). Here the logarithmic spiral provided a better approximation with the resulting pitch angles lying 20-40\,$\degree$ above that of the NIR spiral. The pitch angle is decreasing faster with radius for the logarithmic spirals, with the angles being larger until $\sim$ 126\,au (Sp$_\mathrm{T1}$), 147\,au (Sp$_\mathrm{T1}$) and 106\,au (Sp$_\mathrm{v}$) compared to the linear spiral, yet overall comparable for Sp$_\mathrm{T1}$ and Sp$_\mathrm{v}$ (keeping in mind an uncertainty due to the by-eye parameterization). This is expected, since both functions provide a similar parameterization for these spirals. The largest difference is seen for Sp$_\mathrm{T2}$, where the Archimedean spiral could not sufficiently reproduce the inner parts of the spiral.      
\section{Discussion}\label{sec:Discussion}
Both the gas temperature and the rotation velocity of $^{12}$CO show significant spiral structure over the bulk of the disc when an azimuthally symmetric model is subtracted from the observations (\hyperref[fig:SpiralT]{Figs.~\ref*{fig:SpiralT}},\ref{fig:Spiralv}). Together with a similar feature found in the NIR \citep{Uyama2019} the extent of the structures over a large azimuth ($>$ 180\,$\degree$ temperature, $>$ 360\,$\degree$ velocity) and radius (10-180\,au temperature, 40-180\,au velocity) with a SNR of 2-6 suggests them to be real features. Both the large temperature spiral SP$_{\mathrm{T2}}$ and the NIR spiral appear to follow a similar course, strongly suggesting a link between the two. The location of the spirals matches those of the prominent distortions and bendings occurring in the velocity field and are possibly the cause for the latter. Higher spatial and spectral resolution will be necessary to resolve the very inner disc regions and confirm the presence of the spirals at a higher SNR.

Even though no significant spiral structure is observed in the $^{13}$CO and C$^{18}$O lines, indications for such features are present in both the brightness temperature and rotation velocity residuals of $^{13}$CO and may be made accessible with deeper high resolution observations.

All three spirals observed in $^{12}$CO are well described by a modified logarithmic spiral as well as two  (except SP$_{\mathrm{T2}}$) by an Archimedian (linear) spiral with radially decreasing pitch angles (\hyperref[fig:pitchangles]{Fig.~\ref*{fig:pitchangles}}). The angles found for the logarithmic parameterization mostly lie above those of a linear one. Overall all three spirals are loosely wound and consequently show relatively large pitch angles. These characteristics may be explained through Lindblad-resonance driven spiral wakes of a massive embedded companion (e.g. \citealp{Ogilvie2002,Rafikov2002,Bae2018a,Bae2018b}). 

Since the optically thick $^{12}$CO is tracing higher disc layers, significantly smaller pitch angles, and thus more tightly wound spirals, are expected at the midplane if the disc is passively heated with a positive vertical temperature gradient (compare \citealp{Juhasz2018}). In this context, to distinguish between different spiral launching scenarios it is crucial to use observations spanning the full vertical extent. If the spirals were for example caused by gravitational instability, similar pitch angles are expected for the midplane and the surface layers since the midplane will be heated by shocks in that case. Given the small number of spirals, gravitational instability however seems a rather unlikely cause for the spiral structure we observe (e.g. \citealp{Cossins2009,Hall2020}). On the other hand, several spiral arms could possibly appear as only two or three arms due to resolution effects \citep{Dipierro2014} and the disc around CQ Tau happens to be relatively massive, thus gravitational instability cannot be ruled out at this point. 

So far no clear spiral structures have been found in the dust continuum (or optically thin lines) of CQ Tau to further distinguish possible launching scenarios but could be made accessible with higher spectral and spatial resolution data. Spiral arms have been observed with ALMA in the continuum of several discs, including for example Elias 2-27 \citep{Perez2016}, IM Lup and WaOph 6 \citep{Huang2018}, G17.64+0.16 \citep{Maud2019}, MWC 758 \citep{Boehler2018,Dong2018} and HD100453 \citep{Rosotti2020a}. For the latter, counterparts to the observed NIR spirals \citep{Wagner2015,Benisty2017} were not only found in the dust continuum but also the CO emission, enabling the authors to study the thermal structure of the disc and link the spirals to a known binary companion. The velocity residuals of $^{13}$CO possibly suggest a spiral slightly more tightly wound than the corresponding spiral in $^{12}$CO. A more tightly wound spiral would show smaller pitch angles, which would support \cite{Juhasz2018}.

It is difficult to determine whether the spirals observed in the temperature and the velocity are tracing the same underlying perturbation. Even though the spirals Sp$_{\mathrm{T2}}$ and Sp$_{\mathrm{v}}$ do not fully overlap, they appear to align in the outer parts of the disc between $\sim$ 130-180\,au and 0 to -180\,\degree, hinting towards the same formation mechanism. On the other hand, the calculated pitch angles for the according spirals differ by several degrees. We note however, that the pitch angles are only a rough estimate, since no actual fit was performed and that the actual pitch angles may lie much closer for Sp$_{\mathrm{T2}}$ and Sp$_{\mathrm{v}}$.  

Similarly \cite{Teague2019Spiral} observe temperature and velocity spirals in TW Hydra, that appear to align but do not fully overlap, thus there may be a physical mechanism behind these differences. The authors suggest that layers with different thermal properties are traced, arguing that close to the discs surface spirals in the velocity should be more pronounced due to efficient cooling, while the heat produced by spirals would be more efficiently trapped closer to the midplane. In the case of a companion, the spiral density waves created by either a planet or binary companion will lead to an increase of surface density and thus in a higher CO opacity. This will move the $\tau =1$ layer to a higher altitude, where the temperature in generally higher, resulting in the observed spiral substructure in the gas temperature \citep{Phuong2020b,Phuong2020a}.   

Even though it is impossible to fully disentangle all three velocity components ($v_{r}$, $v_{\mathrm{z}}$, $v_{\phi}$), the same orientation in the residuals as well as the full azimuthal coverage hint towards a vertical perturbation (compare Appendix B in \citealp{Teague2019Spiral}). This is consistent with the (potentially) companion launched spirals in HD 100453 \citep{Rosotti2020a}. As shown by \cite{Pinte2019} an embedded planet will cause perturbations in all three velocity directions, with their strength decreasing with height above the midplane for radial and rotational motions, whereas they increase with height for vertical motions. Gas flowing towards the midplane and falling into the observed cavity could be another explanation for the vertical motions. Since the vertically moving material will receive more stellar light it may appear brighter regardless of the underlying surface density.     

In case the spirals are indeed launched by an embedded planet outside of its orbit, they are expected to converge towards the planet location \citep{Juhasz2015,Bae2018a,Bae2018b}. Since this results in a rapid increase of the pitch angle towards the planet, a possible companion is expected to be located inside of $\sim$ 25\,au in our case. This is consistent with the findings of \cite{Gabellini2019}, who propose an unseen planet of $6-9\,M_{\mathrm{Jup}}$ to explain the deep dust and gas cavity. Given such a location and mass, a companion is not expected to be noticeably affected by extinction in any band \citep{Sanchis2020}. We note, that dynamically launched spirals tend to open up only close to the planet consequently becoming more tightly wound at larger distances, resulting in small pitch angles. It is thus puzzling, that the spirals we observe are still very open far from the possible companion.

In the channel maps presented in \hyperref[fig:channels12CO]{Figs.~\ref*{fig:channels12CO}}, \ref{fig:channels13CO} and \ref{fig:channelsC18O} no clear kinks similar to \cite{Pinte2018,Pinte2019} are detected and the two sides of the disc cannot be resolved due to spectral and spatial resolution limitations. Since several studies \citep{Uyama2019,Gabellini2019} including ours on the other hand indicate the presence of a massive companion of CQ Tau at $< 25$\,au, imprints on the iso-velocities are expected and may be made accessible with higher resolution or sensitivity. 
\section{Summary}\label{sec:Summary}
In this work we presented high angular resolution ALMA observations of $^{12}$CO, $^{13}$CO and C$^{18}$O $J = 2-1$ data of the disc around CQ Tau and used the $^{12}$CO data to analyze the gas temperature and kinematics of the disc. The main results of this analysis are summarised in the following. 
\newline
\newline
\noindent The morphology of the significant spiral structure observed in the brightness temperature and rotation velocity of $^{12}$CO together with the number of spirals and large pitch angles supports a dynamical launching scenario, for example an embedded planet or binary rather than gravitational instabilities. Such a companion is expected to be relatively massive and to be located inside of $\sim$ 25\,au, which is in agreement with \cite{Gabellini2019}. Further multi-line observations at a higher velocity resolution as well as 3D modelling are required to further distinguish the different mechanisms.     

In addition to the gas cavity an intensity drop can be seen in the north-west and south-east side of the disc in all three isotopologues. Postprocessing the DALI model presented by \cite{Gabellini2019} revealed the under-brightness to be caused by beam dilution, rather than by a temperature or line width effect. Since the dip in the south-east appears to be more pronounced, and is co-located with the under-brightness in scattered light, some additional shadowing may still occur, potentially caused by the spiral itself or misaligned regions in the disc.

\begin{figure}[h!]
\centering
    \includegraphics[width=0.49\textwidth]{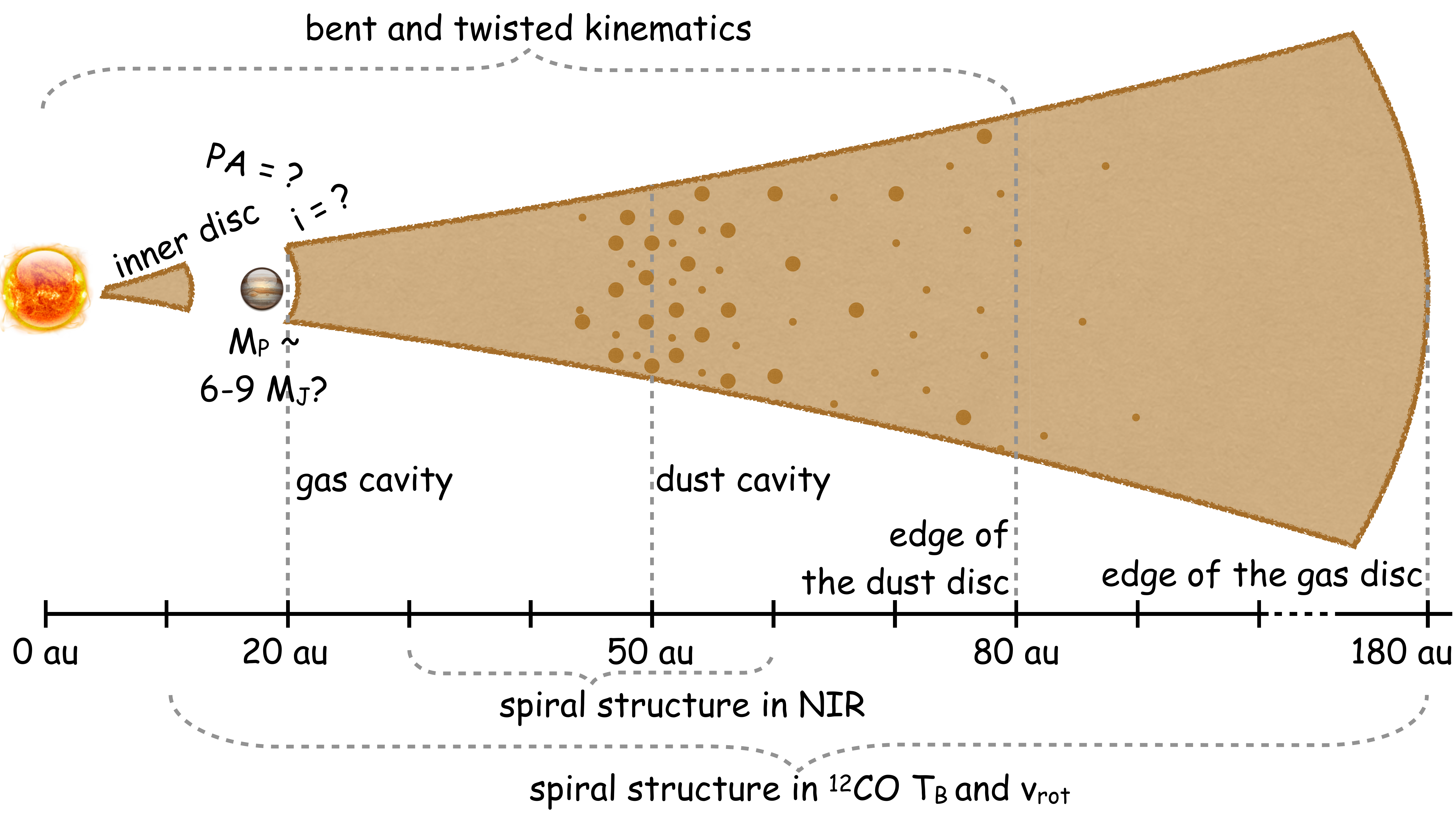}
   \caption{Illustration of the possible morphology of the disc around CQ Tau.}\label{fig:sketch}
\end{figure}

\hyperref[fig:sketch]{Figure~\ref*{fig:sketch}} presents an illustration of the possible morphology of the disc around CQ Tau, taking into account the results from \cite{Gabellini2019}, \cite{Uyama2019} and this work: a dust and gas cavity are present in the disc at $\sim 50$ and $\sim 20$\,au respectively which could be explained by a massive companion inside of 20\,au. Spiral structure is found in both $T_{\mathrm{B}}$, $v_{\mathrm{rot}}$ and the NIR, further supporting the hypothesis of an unseen companion. The inner disc regions, including their position angle and inclination, remain unresolved.
\\
\\
Altogether it appears that the disc around CQ Tau is far from a Keplerian disc. Thus a more detailed non-Keplerian model is required to describe the gas rotation and could be addressed in a future work. Especially the possibility of a massive companion, either a binary star or very massive planet, needs to be further explored. To construct such a model higher angular and spectral resolution ALMA data are essential. Furthermore, NIR interferometry observations at milliarcsec resolution with VLTI-Gravity \citep{gravity} could help to constrain the inclination and position angle of the innermost disc, providing information on the presence of misaligned regions. Combining dust and gas observations of different molecular lines with high SNR is needed to constrain the vertical temperature profile and to further analyse the clearly observed spiral structure of the gas disc.  
%
%

%
\section*{Acknowledgements}
We thank the referee Simon Casassus for his constructive comments and suggestions that helped to improve our work. This paper makes use of the following ALMA data: ADS/JAO.ALMA\#2013.1.00498.S, ADS/JAO.ALMA\#2016.A.000026.S, and ADS/JAO.ALMA\#2017.1.01404.S. ALMA is a partnership of ESO (representing its member states), NSF (USA) and NINS (Japan), together with NRC (Canada) and NSC and ASIAA (Taiwan) and KASI (Republic of Korea), in cooperation with the Republic of Chile. The Joint ALMA Observatory is operated by ESO, auI/NRAO and NAOJ. We acknowledge the support from the DFG Research Unit "Planet Formation Witnesses and Probes: Transition Discs" (FOR 2634/1, ER 685/8-1 $\&$ ER 685/11-1). N.K. acknowledges support provided by the Alexander von Humboldt Foundation in the framework of the Sofja Kovalevskaja Award endowed by the Federal Ministry of Education and Research. GR acknowledges support
from the Netherlands Organisation for Scientific Research (NWO, program number 016.Veni.192.233). This work was partly supported by the Italian Ministero dell’Istruzione, Universit\`{a} e Ricerca (MIUR) through the grant Progetti Premiali 2012 iALMA (CUP C52I13000140001), by the Deutsche Forschungsgemeinschaft (DFG, German Research Foundation) ref. No. FOR 2634/1 TE 1024/1-1, and by the DFG cluster of excellence Origin and Structure of the Universe (www.universe-cluster.de), and by the EC Horizon 2020 research and innovation programme, Marie Sklodowska-Curie grant agreement NO 823823 (Dustbusters RISE project).
%
\bibliographystyle{aa}
\bibliography{bibliography}
\begin{appendix} 
\onecolumn
\section{ALMA observing log}
\begin{table*}[h!]
\centering
\caption{Observing log (ALMA) \label{tab:obs_log}}
\begin{tabular}{|c|c|c|c|c|c|c|}
\hline
Obs ID &    Project   code  & Date       & Configuration &Baselines [m]
& N$_\text{ant}$ & Exp. time [min]  \\
\hline
    1  & 2013.1.00498.S & 30 Aug 2015  & C34-6         & 20 - 1091 &
35             & 15.12 \\
    2  & 2016.A.00026.S & 07 Aug 2017  & C40-7         & 81 - 3700 &
40             & 19.66 \\
    3  & 2017.1.01404.S & 20 Nov 2017  & C43-8         & 92 - 8500 &
44             & 28.73 \\
       &                & 23 Nov 2017  & C43-8         & 92 - 8500 &
48             & 28.73 \\
\hline
\end{tabular}
\end{table*}
\section{Channel maps}\label{appendix:channels}
\begin{figure*}[h!]
    \includegraphics[width=\textwidth]{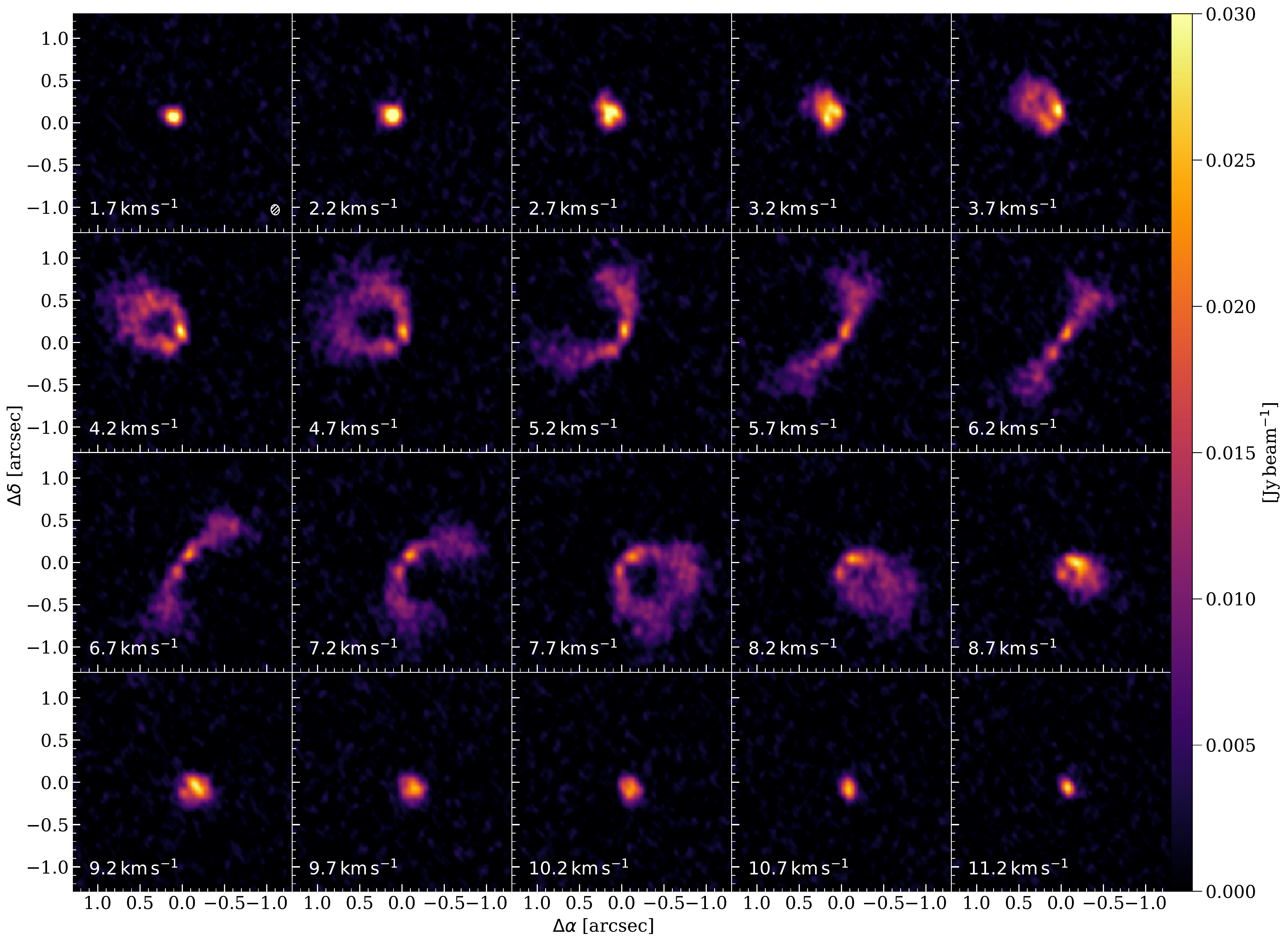}
    \caption{The $^{12}$CO $J=2-1$ line imaged with a channel width of $\Delta v = 0.5\,\mathrm{km}\,\mathrm{s}^{-1}$ and a beam of 0.121x0.098$''$.}
    \label{fig:channels12CO}
\end{figure*}
\begin{figure*}[h!]
    \includegraphics[width=\textwidth]{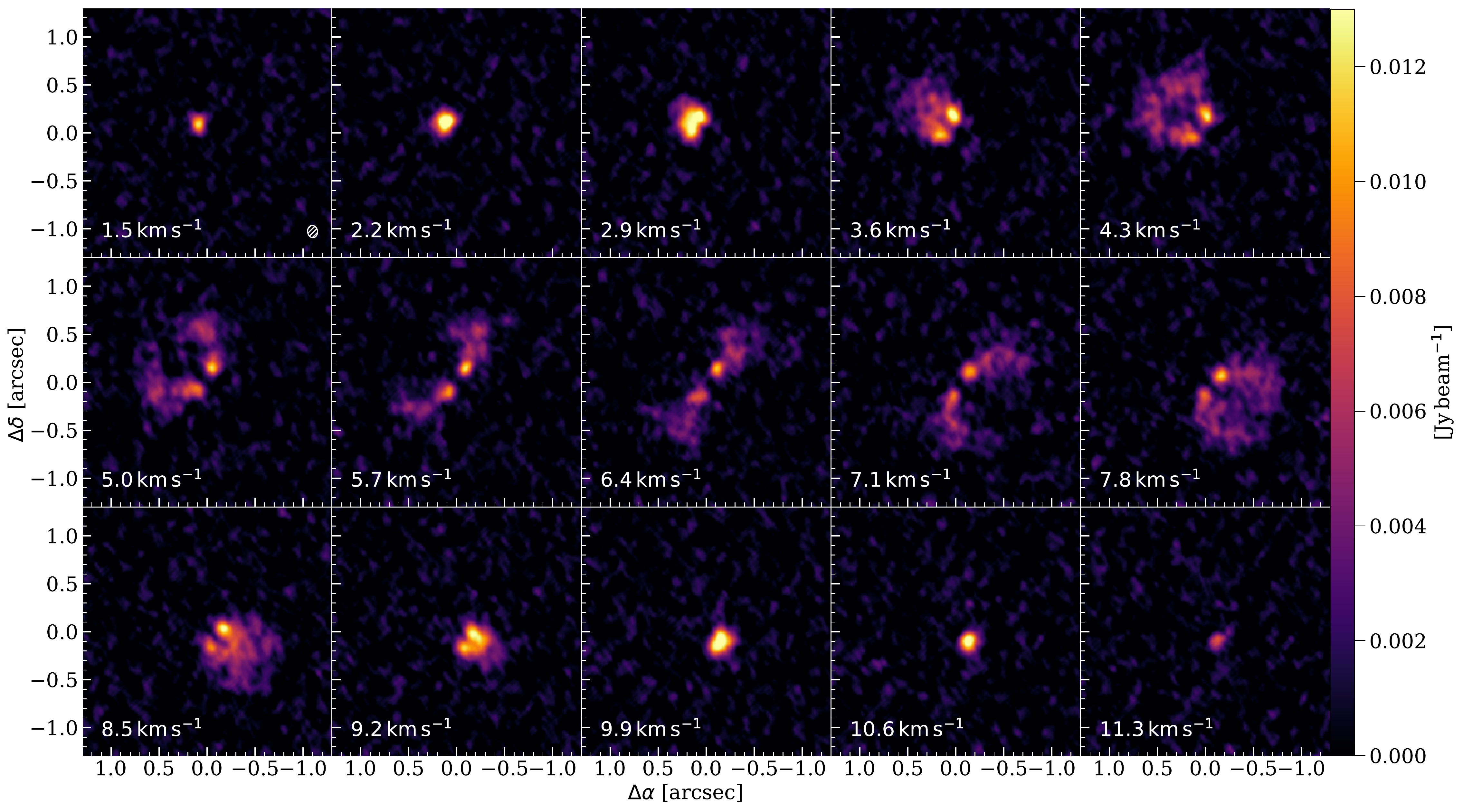}
    \caption{The $^{13}$CO $J=2-1$ line imaged with a channel width of $\Delta v = 0.7\,\mathrm{km}\,\mathrm{s}^{-1}$ and a beam of 0.128x0.103$''$.}
    \label{fig:channels13CO}
\end{figure*}
\begin{figure*}[h!]
    \includegraphics[width=\textwidth]{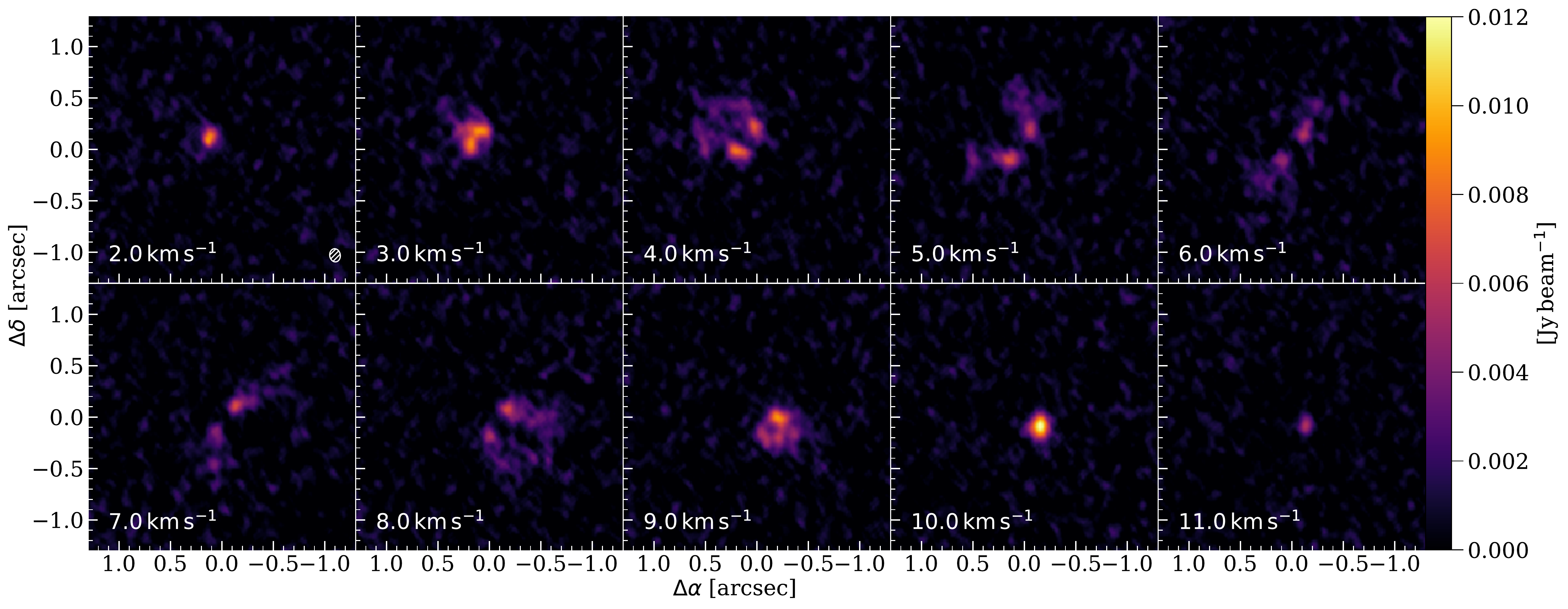}
    \caption{The C$^{18}$O $J=2-1$ line imaged with a channel width of $\Delta v = 1.0\,\mathrm{km}\,\mathrm{s}^{-1}$ and a beam of 0.129x0.103$''$.}
    \label{fig:channelsC18O}
\end{figure*}
\newpage
\section{Beam dilution}\label{appendix:dilution}
\begin{figure*}[h!]
    \includegraphics[width=\textwidth]{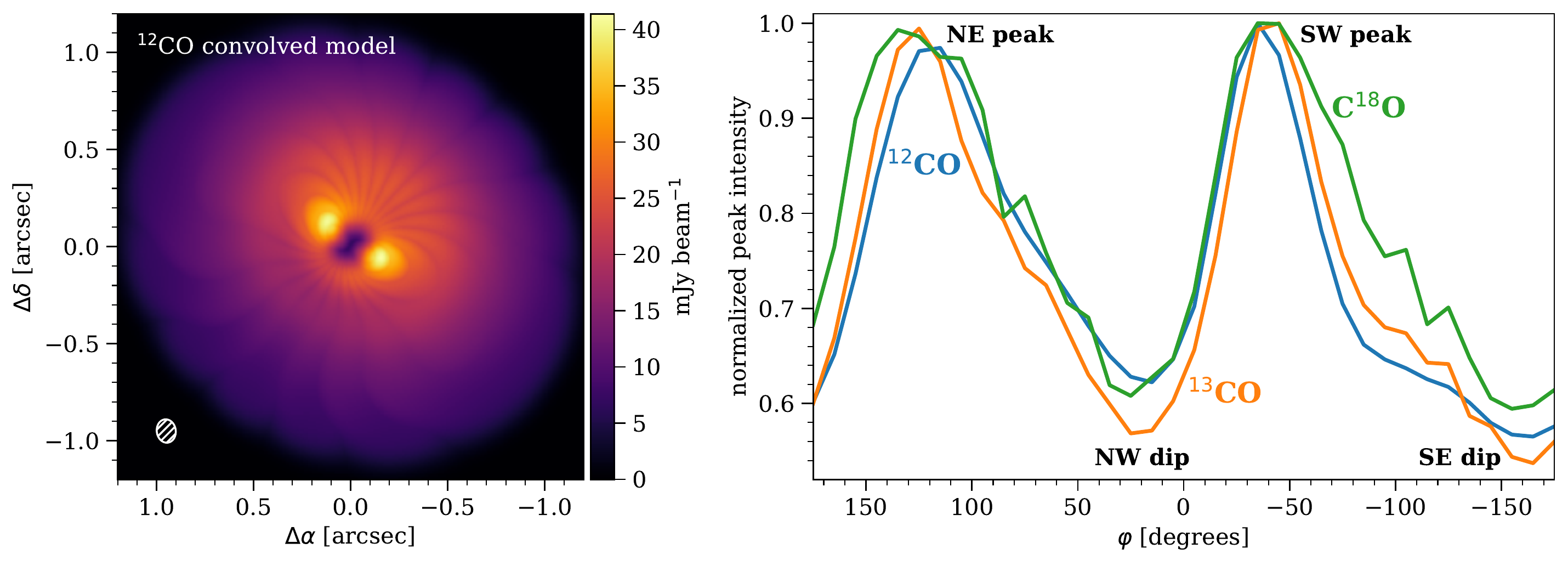}
    \caption{Left: Peak intensity map from postprocessing the DALI model presented by \cite{Gabellini2019}, shown for $^{12}$CO. Right: Azimuthal variations of the peak intensity of $^{12}$CO, $^{13}$CO and C$^{18}$O for an annulus of 20-40\,au, derived from the DALI model.}
    \label{fig:dilution}
\end{figure*}
\newpage
\section{Posterior distributions of the modelling}\label{appendix:table}
%
%
\begin{table*}[h!]
\centering
\caption{Posterior distributions of the razor-thin disc models. The uncertainties represent the 16th to 84th percentiles about the median value.}\label{tab:ParametersThin}
\centering
\begin{tabular}{l cccccccc}
\hline
\hline
run & $r_{\mathrm{in}}$ [$''$] & $r_{\mathrm{out}}$ [$''$] & $x_0 $ [$''$] & $y_0$ [$''$] & PA [$\degree$]& $v_{\mathrm{LSR}}$ [m\,s$^{-1}$] & $M_*$ [M$_{\odot}$] \\
\hline
\multicolumn{8}{l}{\textit{n$_{\mathit{walkers}}$ = 200, n$_{\mathit{burnin}}$ = 5000, n$_{\mathit{steps}}$ = 5000}} \\
\multicolumn{8}{l}{\textbf{Full disc}} \\
\hline
\vspace{0.1cm}
1 & 0.0 & 1.0 & 0.0089 $\pm$ 0.0002& 0.0064 $\pm$ 0.0001 & 234.22 $\pm$ 0.03& 6177.1 $\pm$ 0.8& 1.582 $\pm$ 0.002 \\
\vspace{0.1cm}
2 & 0.25 & 1.0 & 0.0072 $\pm$ 0.0003& 0.0273 $\pm$ 0.0003 & 234.58 $\pm$ 0.03& 6172.9 $\pm$ 0.8& 1.570 $\pm$ 0.002 \\
\hline
\multicolumn{8}{l}{\textbf{Inner disc}} \\
\hline
\vspace{0.1cm}
3 & 0.0 & 0.4 & 0.0116 $\pm$ 0.0002 & 0.0051 $\pm$ 0.0001 & 232.98 $\pm$ 0.07 & 6124.1 $\pm$ 2.3& 1.626 $\pm$ 0.003\\
\vspace{0.1cm}
3 & 0.0 & 0.5 & 0.0106 $\pm$ 0.0002& 0.0058 $\pm$ 0.0001 & 233.11 $\pm$ 0.05 & 6145.8 $\pm$ 1.6& 1.642 $\pm$ 0.002\\
\hline
\multicolumn{8}{l}{\textbf{Outer disc}} \\
\hline
\vspace{0.1cm}
5 & 0.4 & 1.0 & 0.0066 $\pm$ 0.0004& 0.0130 $\pm$ 0.0004 & 234.69 $\pm$ 0.04& 6183.7 $\pm$ 0.9 & 1.570 $\pm$ 0.002 \\
\vspace{0.1cm}
6 & 0.5 & 1.0 & 0.0061 $\pm$ 0.0005& 0.0114 $\pm$ 0.0006 & 234.87 $\pm$ 0.04 & 6185.6 $\pm$ 1.0 & 1.541 $\pm$ 0.002 \\
\hline
\multicolumn{8}{l}{\textbf{Annuli}} \\
\hline
\vspace{0.1cm}
7 & 0.2 & 0.4 & 0.0124 $\pm$ 0.0004& 0.0282 $\pm$ 0.0005& 233.57 $\pm$ 0.08& 6095.8 $\pm$ 2.5& 1.580 $\pm$ 0.003 \\
\vspace{0.1cm}
8 & 0.4 & 0.6 & 0.0033 $\pm$ 0.0006& 0.0141 $\pm$ 0.0006& 233.67 $\pm$ 0.06& 6179.9 $\pm$ 1.7& 1.636 $\pm$ 0.003 \\
\vspace{0.1cm}
9 & 0.6 & 0.8 & 0.0031 $\pm$ 0.0008& 0.0245 $\pm$ 0.0008& 235.60 $\pm$ 0.06& 6169.4 $\pm$ 1.4& 1.473 $\pm$ 0.003 \\
\vspace{0.1cm}
10 & 0.8 & 1.0 & 0.0242 $\pm$ 0.0013& 0.0020 $\pm$ 0.0012& 235.33 $\pm$ 0.07& 6194.9 $\pm$ 1.7& 1.650 $\pm$ 0.004 \\
\hline
\hline
\end{tabular}
\end{table*}
\newpage
\end{appendix}
\end{document}